\documentclass[12pt,preprint]{aastex}
\begin{document}
\bibliographystyle{apj}

\title{A Probabilistic Approach to Classifying Supernovae Using Photometric Information}

\author{
Natalia V. Kuznetsova\altaffilmark{1},
Brian M. Connolly\altaffilmark{2}
}
\altaffiltext{1}{Lawrence Berkeley National Lab, Berkeley, CA}
\altaffiltext{2}{Department of Physics, Columbia University, New York, NY}

\email{nvkuznetsova@LBL.gov, connolly@nevis.columbia.edu}


\begin{abstract}
This paper presents a novel method for determining the probability
that a supernova candidate belongs to a known supernova type (such as 
Ia, Ibc, IIL, \emph{etc.}), using its photometric information alone. 
It is validated with Monte Carlo, and both space- and ground- based data.  
We examine the application of the method to well-sampled as 
well as poorly sampled supernova light curves and investigate to what
extent the best currently available supernova models can be used 
for typing supernova candidates.  
Central to the method is the assumption that a supernova candidate
belongs to a group of objects that can be modeled; we therefore discuss 
possible ways of removing anomalous or less well understood events
from the sample.  This method is particularly advantageous for
analyses where the purity of the supernova sample is of the essence,
or for those where it is important to know the number of the
supernova candidates of a certain type (\emph{e.g.}, in 
supernova rate studies).  

%It is advantageous to methods involving $\chi^2$'s because it 
%accounts for the alternative hypotheses for the candidate.  
%Also, we suggest a maximum likelihood method to fit supernova parameters.  

\end{abstract}

\keywords{supernovae: general}
%\keywords{supernovae -- cosmology -- Type Ia supernovae}

%\maketitle

%%%%%%%%%%%%%%%%%%%%%%%%%%%%%%%%%%%%%%%%%%%%%
\section{Introduction}
%%%%%%%%%%%%%%%%%%%%%%%%%%%%%%%%%%%%%%%%%%%%%
Type Ia supernovae, empirically established to be standardized candles, 
are a staple of experimental cosmology.  A number of future
cosmological probes (\emph{e.g.}, DES, Pan-STARRS, LSST, JDEM) 
are planning massive sky surveys that will collect very 
large samples of supernovae, which will have to be classified into several
known types.  
Supernova candidates are most easily classified by their spectra,
but for very large ground-based surveys it may be  impractical to obtain
a spectral confirmation of the supernova type for each candidate.  Therefore, it is 
 imperative to develop reliable methods of supernova classification
based on photometric information alone.  Photometric typing has 
been described in~\cite{pozn02},~\cite{riess2004},~\cite{john} and
~\cite{sull1}, among others.
Most of the existing methods rely on color-color or color-magnitude 
diagrams for supernova classification.

In this paper, we propose a novel approach to the photometric typing of 
supernova candidates.  This method is based on a probability derived 
using a Bayesian approach, and is well suited for extracting the maximum amount 
of information out of limited data.  Unlike methods relying on 
color-color or color-magnitude diagrams, which require a comparison
of the candidate's color(s) with pre-existing tables or plots -- a
comparison that requires a good understanding of the errors
and assumptions that went into the making of the literature data -- 
our approach calls for a calculation of a single number 
and automatically takes into account all of the 
information that is available for a given candidate while incorporating
the currently best-known models for supernova behavior.

A Bayesian method in the context of supernova light curve fitting
has been used in~\cite{barris}; however, it was applied specifically to
type Ia supernovae to deduce redshift-independent distance moduli.   
In this paper, on the other hand, we describe
a probabilistic approach to typing photometrically surveyed supernovae.
Such an approach allows the possibility of the marginalization
(integration) of the unknown, or nuisance, parameters.
In contrast to a traditional $\chi^2$ calculation, this technique is simply a 
calculation of a probability and does not involve fitting or minimization.
However, it does require that the candidate sample be well understood;
or, in other words, that each candidate be one of a number of hypothesized
objects whose behavior can be well modeled.

The purpose of this paper is twofold.  Apart from introducing a methodology
that can be easily applied ``as is'' or extended as needed, we would like to
test the extent to which applying the best currently available supernova models
helps in typing supernova candidates.  There is no doubt that within the next
several years new and improved supernova models will be constructed; when they are, they 
can be easily worked into the method.

The paper is structured as follows.  In Section~\ref{sec:like},
we derive the probability that 
a given candidate is a Type Ia supernova.  In Section~\ref{sec:valid}, 
we discuss how well the method works when applied to poorly sampled,
Hubble Space Telescope GOODS data, and to well-sampled, ground-based SNLS data.
We suggest further improvements to the method in Section~\ref{sec:improve},
discuss its possible application to ``anomalous'' objects 
in Section~\ref{sec:anomalies}; and to fitting for supernova 
parameters, in Section~\ref{sec:fitting}.
Conclusions are given in Section~\ref{sec:concl}.

%%%%%%%%%%%%%%%%%%%%%%%%%%%%%%%%%%%%%%%%%%%
\section{The Probability}
\label{sec:like}
%%%%%%%%%%%%%%%%%%%%%%%%%%%%%%%%%%%%%%%%%%%
Let us suppose that we have a sample of supernova candidates
where it has been established that every candidate is 
consistent with some type of astronomical object with 
known or well-modeled
photometric behavior 
(one approach to making sure that this is indeed the case
is discussed in Section~\ref{sec:anomalies}).
In our example, we consider photometric models for type 
Ibc IIL, IIP, IIn, and standard, or ``Branch-normal''~\citep{branch}, type Ia
supernovae because they are currently best known;
however, the method can be trivially 
extended to include other supernova types, as well as variable
objects that are not supernovae, once reliable models for such objects are available.
We would like to determine the probability that a
given candidate in the sample is a supernova of some
known type $T$, given its measured light curve data.
Here, we will focus on the case where 
$T$ is Ia, but the method can be easily applied to other types as well.

Here and for the remainder of the paper we  
assume that the redshift of the supernova candidate is perfectly
known.  We  begin by assuming that the light curve of the candidate is
measured in a single broadband filter $a$ (the method can be trivially
extended to any number of filter bands).  The light curve measurements
are represented in $n_{epochs}$ (observing times) 
as {$\{A_i\}$}, where $i$ = [1,...,$n_{\rm epochs}$].  
We would like to know the probability that this candidate
is a type $T$ supernova, $P(T|\{A_i\})$.
Technically, since $\{A_i\}$ defines the candidate, 
we are calculating the probability that the type $T$ hypothesis
is true given a candidate, or:
\begin{equation}
P(T|\{A_i\})\equiv P(T|{\rm candidate}).
\end{equation}
The probability $P(T|\{A_i\})$ 
depends on many parameters. 
We will express $P(T|\{A_i\})$ as a function of 
these parameters and, in the end, marginalize them to quantify
$P(T|\{A_i\})$ .

Assume now that we have a photometric model (which we will 
also refer to as a template), $\{a_j\}$, for 
the expected light curve for a supernova of type $T$ at a given redshift, 
observed in filter $a$.
In general, this model depends on parameters such as the date
of maximum light $t_0$ (or, equivalently, the time difference between 
the dates of maximum light for the model and the data, $t_{\rm diff}$
%, such that $j \equiv i + t_{\rm diff}$
);
stretch $s$~\citep{perl}, which parametrizes the width of the 
light curve (if $T$ = Ia);
the assumed absolute  magnitude $M$ in the restframe $B$-band;
and interstellar extinction parametrized, \emph{e.g.}, 
by the Cardelli-Clayton-Mathis parameters $R_v$ and $A_v$~\citep{ccm}.
We will refer to this collection of parameters as
\begin{equation}
\vec{\theta} \equiv (t_{\rm diff},s,M,R_v,A_v).
\end{equation}
The parameters $\vec{\theta}$ uniquely 
define $\{a_j\}$ for type $T$:
\begin{equation}
\{a_j\}\rightarrow\{a_j(\vec{\theta},T)\}
\end{equation}

In this study, we would like to determine the probability $P({\rm Ia}| \{A_i\})$ that the candidate 
is a type Ia supernova given the measurements $\{A_i\}$.
Unfortunately, this probability is not directly calculable, but one can easily calculate
$P(\{A_i\}|{\rm Ia} )$, the probability of obtaining the measurement given that the candidate
is a type Ia supernova.
Bayes' theorem allows us to relate these two quantities.  Na\"{\i}vely,
we may write
\begin{equation}
P({\rm Ia}| \{A_i\}) = \frac{P(\{A_i\} | {\rm Ia}) \, P({\rm Ia})} {\sum_{T} P(\{A_i\} | T) \, P(T) }
\end{equation}
where $P(\{A_i\} | Ia)$ is the probability to obtain  data $\{A_i\}$
for supernova type Ia, $P({\rm Ia})$ contains prior information about type Ia supernovae,
and the denominator is the normalization over all of the known supernova types $T$.
This sum is over a finite number of supernova types, which gives a legitimate 
probability because
we assume that each candidate must be one of the types summed in the denominator.
Of course, one would like the denominator to include a model 
for every possible object that could mimic or be a supernova; in practice, 
we have to limit ourselves to a subset of major supernova types with well-known
photometric models.

When calculating $P({\rm Ia}|\{A_i\})$ one must consider that a range of possible 
values is allowed for the stretch, extinction, and other parameters
that characterize a Type Ia supernova.  
Therefore, we express $P({\rm Ia}|\{A_i\})$ as a function of these
parameters and then marginalize them:
\begin{eqnarray}
P({\rm Ia}|\{A_i\})=\sum_{\vec{\theta}} \, P(\vec{\theta}, {\rm Ia}|\{A_i\}) \nonumber \\
= \frac{\sum_{\vec{\theta}} \, P(\{A_i\}|\vec{\theta},{\rm Ia})P(\vec{\theta},{\rm Ia})}{\sum_{T} \, \sum_{\vec{\theta}} \, P(\{A_i\}|\vec{\theta},T)P(\vec{\theta},T)}.
\label{eqn:bayes}
\end{eqnarray}
%\begin{eqnarray}
%\Delta\vec{\theta} \equiv \Delta t_{diff} \Delta M \Delta s 
%\label{eqn:bayes}
%\end{eqnarray}
The prior probability $P(\vec{\theta}, T)$ contains all known information about 
the distributions of $t_{\rm diff}$, $M$, $s$, and other relevant parameters for type $T$
supernovae.  Note that 
$P(\vec{\theta},T)$ could also 
include the relative probabilities of obtaining a type $T$ supernova,
\emph{i.e.}, the relative rates of the different supernova types.
However, in this study we will assume no prior
knowledge of the relative supernova 
rates, and instead assume the rates are all the same.

Because $\vec{\theta}$ and $T$ uniquely define $\{a_j\}$,
$P(\{A_i\} | \vec{\theta}, T)$ can be written as:
\begin{eqnarray}
P(\{A_i\} |\vec{\theta}, T) \equiv P(\{A_i\} |\{a_j\}).
\end{eqnarray}
The measured light curve, $\{A_i\}$, is taken to be in units
of counts/second, and can fluctuate from the model $\{a_j\}$ according to Gaussian statistics:
\begin{equation}
\label{equation:likelihood}
P(\{A_i\} |\vec{\theta}, T)  \equiv P(\{A_i\} |\{a_j\} = \prod_{i=1}^{n_{epochs}} \frac{e^{-\frac{(a_j-A_i)^2}{2 \delta A_i^2}} }{\sqrt{2\pi} \delta A_i} \Delta a_j .  
\end{equation}
In this expression, $A_i$ and $\delta A_i$ are experimental measurements and errors for epoch $i$,
 $\{a_j\}$ represents the predicted light curve for a type $T$ supernova,
and $\Delta a_j$ is an  increment of $a_j$.

We use the models (represented by $\{a_j\}$) for supernova Ia, Ibc, IIL, IIP, and IIn from~\cite{nugent}.
Several issues are of the essence here.
First, the model for type Ia supernovae extends both into the UV (below 3460 \AA\, in the supernova
rest frame) and into the IR (above 6600 \AA\, in the supernova rest frame) regions.  Both of these
regions are poorly constrained with the currently available data.  This is in fact the reason why 
some authors~\citep[for example]{salt} choose
to limit themselves to only the well-known part of the type Ia supernova spectrum.  Second,
the behavior of non-type Ia supernovae, especially at high redshifts, is not well known; 
the currently available models may or may not be an adequate representation 
of these objects.
In our work, however, we are interested in exploring the extent to which the best possible 
supernova models currently available offer a discriminating power.  A Bayesian
approach that allows one to easily work in the uncertainties on any prior
knowledge of supernovae appears quite natural for the situation.  
The supernova models used in this work are discussed in more detail in Appendix A.

In order to calculate the prior $P(\vec{\theta}| T)$,
we make the simplifying assumption that $R_v$ and $A_v$,
$t_{\rm diff}$, $M$ and $s$ are all independent, 
allowing us to factorize the probability:
\begin{equation}
P(\vec{\theta} |T ) = P(t_{\rm diff}|T) \, P(M|T)\, P(s|T)\, P(R_v,A_v|T)\, P(T).
\label{equation:prior}
\end{equation}

Let us consider each of the terms in Eqn.~\ref{equation:prior} in turn,
starting with the probability $P(T)$ of observing a supernova
of type $T$.
Since we assume no prior knowledge of the relative rates of the various 
types, each type has an equal probability of appearing
in the candidate sample.  Therefore, $P(T)$ is a constant:
\begin{equation}
P(T)\equiv \frac{1}{N_T},
\end{equation}  
where $N_T$ is the number of supernova types considered.

We also assume a flat prior for the difference in the dates of maximum light between
the model and the data, $t_{\rm diff}$.  
In practice, we compare the measured and modeled light curves shifting
their relative dates of maximum by one day.
Marginalization over $t_{\rm diff}$  thus means
summing over a finite number of such shifts.  
As each shift has an equal probability and the prior must be normalized to 1,
\begin{equation}
P(t_{\rm diff}|T) \equiv \frac{1}{N_{\rm diff}} = \frac{\Delta t_{\rm diff}}{{t^{max}_{\rm diff}-t^{min}_{\rm diff}}}.
\end{equation}  
where $\Delta t_{\rm diff}$ is an  increment of $t_{\rm diff}$, 
$N_{\rm diff}$ is the number of shifts, 
and the maximum $t^{max}_{\rm diff}$ and minimum $t^{min}_{\rm diff}$
set the limits on $t_{\rm diff}$.    We take $N_{\rm diff}$ to be 160.

There are potentially large uncertainties associated with how well  
the luminosity function of a particular type supernova is known.
This is accounted for in the prior $P(M|T)$:
\begin{equation}
P(M|T) =\frac{e^{-\frac{(M-\bar{M})^2}{2 \delta M^2}}}{\sqrt{2 \pi} \delta M} \Delta M.
\end{equation}  
Here, $\Delta M$ is an  interval in $M$.
We extract the mean magnitudes $\bar{M}$ in the restframe $B$-band from~\cite{nugent},
and the standard deviations, $\delta M$, from~\cite{rich}. 
Table~\ref{tab:mags} summarizes the values used.

\clearpage
%%%%%%%
\begin{table}[htbp]
  \begin{center}
    \begin{tabular}{|c|c|c|}
    \hline\hline
      Supernova Type     & $\bar{M}$   & $\delta M$ \\
     \hline\hline 
       Ia                & -19.05      &  0.30 \\
       Ibc               & -17.27      &  1.30 \\
       IIL               & -17.77      &  0.90 \\
       IIP               & -16.64      &  1.12 \\
       IIn               & -19.05      &  0.92 \\
     \hline\hline
   \end{tabular}
 \end{center}
\caption[]
{\label{tab:mags}
Mean magnitudes (Vega, restframe $B$-band) and standard deviations for different supernova types.}
\end{table}
%%%%%%%
\clearpage
Note that there is an additional dispersion in magnitudes due to lensing effects; see, for 
example,~\cite{ald}.  We will ignore this effect in our study.

Concerning the prior on the stretch $s$, $P(s|T)$,
we make a simplifying assumption that for type Ia supernovae
the distribution for the stretch is the same for all filter bands.
The concept of stretch itself, strictly speaking, is only defined in the restframe $B$-
and $V$- bands, although an extension of the concept into the restframe $I$-band is discussed 
in~\cite{nob}; an implementation of this extension would introduce several extra 
parameters into the likelihood calculation.
For type Ia's, the stretch is known to have a distribution given in~\cite{sull}.  
Approximating this distribution by a Gaussian:
\begin{equation}
P(s|{\rm Ia}) =\frac{e^{-\frac{(s-\bar{s})^2}{2 \delta s^2}}}{\sqrt{2 \pi} \delta s} \Delta s,
\label{eqn:sprioria}
\end{equation}
where $\Delta s$ is an increment in $s$, and 
we obtain $\bar{s}$ = 0.97 and $\delta s$ = 0.09.
For non-type Ia's, the stretch is not defined.  Practically 
speaking, this means that the light curves of non-type Ia candidates 
do not depend on the value of the stretch.  
We therefore assume a flat
prior such that:
\begin{equation}
P(s|T) =\frac{1}{s_{max}-s_{min}} \Delta s \, \, \, ({\rm for}~T\ne {\rm Ia})
\label{eqn:spriornonia}
\end{equation}
where $s_{max}$ and $s_{min}$ can be arbitrary, provided they are consistent with the limits
on the stretch considered for the type Ia case.  Appendix B details 
why it is necessary to retain the stretch parameter for non-type Ia 
supernovae.

Lastly, we parametrize the effects of the interstellar extinction as in~\cite{ccm}.
The model supernova light curves are generated with some particular values of 
the Cardelli-Clayton-Mathis parameters $A_v$ and $R_v$.
Introducing the actual distributions for $A_v$ and $R_v$ is difficult due to lack 
of a generally accepted model.  For example, there have been several indications
that the Milky Way value of $R_v$ = 3.1 may not be generally applicable.  Values
in the range of $\sim$2 to 3.5 have been suggested~\citep{dust1,dust2}.
Likewise, there is no consensus for the distributions of $A_v$'s, although there do exist
studies offering various non-analytic parametrization (for example,~\cite{hatano} introduce
a model of the extinction distribution as a function of the galaxy inclination, for 
both Ia and core-collapse supernovae).
In our study, we compromise by considering a case of no extinction, as well
as two cases of extinction with a moderate value of $A_v$ = 0.4 and two 
different values of $R_v$, 2.1 and 3.1.  All three cases ($N_v = 3$) are considered
equally possible. In other words, we take:
\begin{eqnarray}
P(R_v,A_v|T) = \frac{1}{N_v},
\end{eqnarray}
where $N_v$ are the number of possible discrete values of $R_v$ and $A_v$ considered.  
It is worth noting that once more generally accepted 
models for $A_v$ and $R_v$ appear, it will be trivial to introduce them into the formalism.

%It is now possible
%to calculate $P(\{A_i\}| \{a_j(\vec{\theta}, T)\})\, P(\vec{\theta} | T)$ for both Ia and 
%non-Ia cases; for $T = {\rm Ia}$:
%\begin{small}
%\begin{eqnarray}
%P(\{A_i\}| \vec{\theta}, {\rm Ia}) \, P(\vec{\theta} | {\rm Ia}) = \nonumber \\
%\frac{1}{N_v}\frac{1}{N_T}\frac{1}{N_{t_{\rm diff}}}\frac{\Delta M}{\sqrt{2 \pi} \delta M} \, e^{-\frac{(M-\bar{M})^2}{2 \delta M^2}}\, \frac{\Delta s}{\sqrt{2 \pi} \delta s} \, e^{-\frac{(s-\bar{s})^2}{2 \delta s^2}} \, \prod_{i=1}^{n_{epochs}} \frac{ \Delta a_j}{\sqrt{2 \pi} \delta A_i} \, e^{-\frac{(a_{j}-A_i)^2}{2\delta A_{i}^2}},
%\end{eqnarray}
%\end{small}
%and for $T \ne {\rm Ia}$:
%\begin{small}
%\begin{eqnarray}
%P(\{A_i\}| \vec{\theta},T) \, P(\vec{\theta} | T) = \nonumber \\
%\frac{1}{N_v}\frac{1}{N_T}\frac{1}{N_{t_{\rm diff}}}\frac{\Delta M}{\sqrt{2 \pi} \delta M} \, e^{-\frac{(M-\bar{M})^2}{2 \delta M^2}}\, \frac{\Delta s}{s_{max}-s_{min}} \, \prod_{i=1}^{n_{epochs}} \frac{ \Delta a_j}{\sqrt{2 \pi} \delta A_i} \, e^{-\frac{(a_{j}-A_i)^2}{2\delta A_{i}^2}}.
%\end{eqnarray}
%\end{small}
We can now put everything together to calculate 
$\sum_{\vec{\theta}} \, P(\{A_i\}| \vec{\theta}, T) P(\vec{\theta}, T)$ 
for both Ia and non-Ia cases;  for $T = {\rm Ia}$:
\begin{small}
\begin{eqnarray}
\label{eqn:final_Ia}
\sum_{\vec{\theta}} \, P(\{A_i\}| \vec{\theta}, {\rm Ia}) P(\vec{\theta}, {\rm Ia}) = \nonumber \\
\frac{1}{N_T}\sum_{t_{\rm diff}}\frac{1}{N_{t_{\rm diff}}}\sum_{R_v,A_v}\frac{1}{N_v}\sum_{M=M_{min}}^{M_{max}} \frac{\Delta M}{\sqrt{2 \pi} \delta M} \, e^{-\frac{(M-\bar{M})^2}{2 \delta M^2}}\, \sum_{s=s_{min}}^{s_{max}} \frac{\Delta s}{\sqrt{2 \pi} \delta s} \, e^{-\frac{(s-\bar{s})^2}{2 \delta s^2}} \nonumber \\ \prod_{i=1}^{n_{epochs}} \frac{ \Delta a_j}{\sqrt{2 \pi} \delta A_i} \, e^{-\frac{(a_{j}-A_i)^2}{2\delta A_{i}^2}},
\end{eqnarray}
\end{small}
and for $T \ne {\rm Ia}$:
\begin{small}
\begin{eqnarray}
\label{eqn:final_T}
\sum_{\vec{\theta}} \, P(\{A_i\}| \vec{\theta}, T) P(\vec{\theta}, T) = \nonumber \\
\frac{1}{N_T}\sum_{t_{\rm diff}}\frac{1}{N_{t_{\rm diff}}}\sum_{R_v,A_v}\frac{1}{N_v}\sum_{M=M_{min}}^{M_{max}} \frac{\Delta M}{\sqrt{2 \pi} \delta M} \, e^{-\frac{(M-\bar{M})^2}{2 \delta M^2}}\, \sum_{s=s_{min}}^{s_{max}} \frac{\Delta s}{s_{max}-s_{min}}  \nonumber \\ \prod_{i=1}^{n_{epochs}} \frac{ \Delta a_j}{\sqrt{2 \pi} \delta A_i} \, e^{-\frac{(a_{j}-A_i)^2}{2\delta A_{i}^2}}.
\end{eqnarray}
\end{small}

Finally, we can insert Eqns.\,(\ref{eqn:final_Ia}) and (\ref{eqn:final_T}) into 
(\ref{eqn:bayes}) to obtain $P({\rm Ia}|\{A_i\})$.
Naturally, if there are light curve measurements in more than one bandpass, the 
probability can be calculated for all of the available filters, each with 
its own photometric model dependent on the same parameters as in
the single-filter case.
For multi-filter measurements, there will be as many products over epochs in
Eqns.\,(\ref{eqn:final_Ia}) and (\ref{eqn:final_T}) as there are  
passbands available.

Note that if the redshift of a supernova is not well known, it can also be introduced
as a nuisance parameter.  Likewise, the method can be trivially extended to include any
other parameter of interest.

%%%%%%%%%%%%%%%%%%%%%%%%%%%%%%%%%%%%%%%%%%%
\section{Demonstrating the Method}
\label{sec:valid}
%%%%%%%%%%%%%%%%%%%%%%%%%%%%%%%%%%%%%%%%%%%
In order to demonstrate the method, we consider the following samples:
\begin{itemize}
\item
Poorly sampled Hubble Space Telescope (HST) GOODS data consisting of both type 
Ia's and non-Ia's.  
\item
Monte Carlo events with the same poor sampling as in the GOODS data.  
\item
Well-sampled ground-based data consisting of all type Ia's.
\end{itemize}
In all cases, we used the simulation described in Appendix A to create
template light curves in the filter bands considered.

%%%%%%%%
\subsection{The Space-based GOODS Data}
\label{sec:space}
%%%%%%%%
To illustrate how the method performs on space-based data when measurement epochs are scarce, 
we use the gold (``high confidence'') and silver (``likely but not certain'') 
candidates from the HST GOODS sample~\citep{goods0,goods1,goods2}.  
The gold and silver classification is described in~\cite{riess},
and refers to the level of confidence with which the type of a supernova 
was determined.  The sample includes 15 gold and 5 silver Ia and 1 gold and 6 silver  
core-collapse (CC) candidates.

In addition to the GOODS data, we use a sample collected by the Supernova Cosmology Project (SCP)
and the High-Z collaborations in the spring-summer of 2004.
The latter sample covers the North GOODS 
field only and consists of 4 epochs separated by approximately 45 days.
Because the data were recorded in only two filter passbands, 
HST ACS F775W  and F850LP, in this section
we will restrict ourselves to only using these two bands.
%Each $I$-band image has a single dither; each $Z$-band image, four.   

We start with the data that have been flat-fielded and gain-corrected
by the HST pipeline, and use MultiDrizzle~\citep{drizzle} to 
perform cosmic ray rejection and to combine dithered observations.
%process it making use of the STSDAS,
%PyRAF (\cite{pyraf})\footnote{STSDAS and PyRAF are products of the Space 
%Telescope Science Institute, which is operated by AURA for NASA},
%and MultiDrizzle (\cite{drizzle}) packages.
%The parameters
%of the drizzling process include a  ``square'' kernel, with a pixel fraction of 0.66
%and a pixel scale of 1.0.  The drizzling combines the multiple individual pointings
%to produce an image free of cosmic rays.  Notice, however, that this procedure
%is ineffective for the $I$-band data from the SCP sample, since they contain
%only a single pointing for each tile.  We therefore
%use a morphological cosmic ray rejection package (\cite{lacosmics}) to 
%create images with identifiable objects, thus allowing us to generate 
%the geometrical transformations between images.  Only these transformations 
%are then used for aligning the original images (contaminated by cosmic rays) with
%other images; the original images are used to extract photometry.
%After the images are thus cleaned of cosmic rays, 
We search for and perform simple aperture photometry on the supernova candidates 
in each of the five GOODS epochs.  To obtain the multi-epoch photometry
for the GOODS North data, we combine all four epochs of the SCP/High-Z sample
and then subtract these data from each of the North GOODS epochs in turn.
For the GOODS South data, we combine South epochs 4 and 5, as well as 1 and 2;
we then subtract the two combined samples separately from each of the five 
South GOODS epochs.

\begin{figure}[htb]
 \begin{center}
    \includegraphics[width=5.0in]{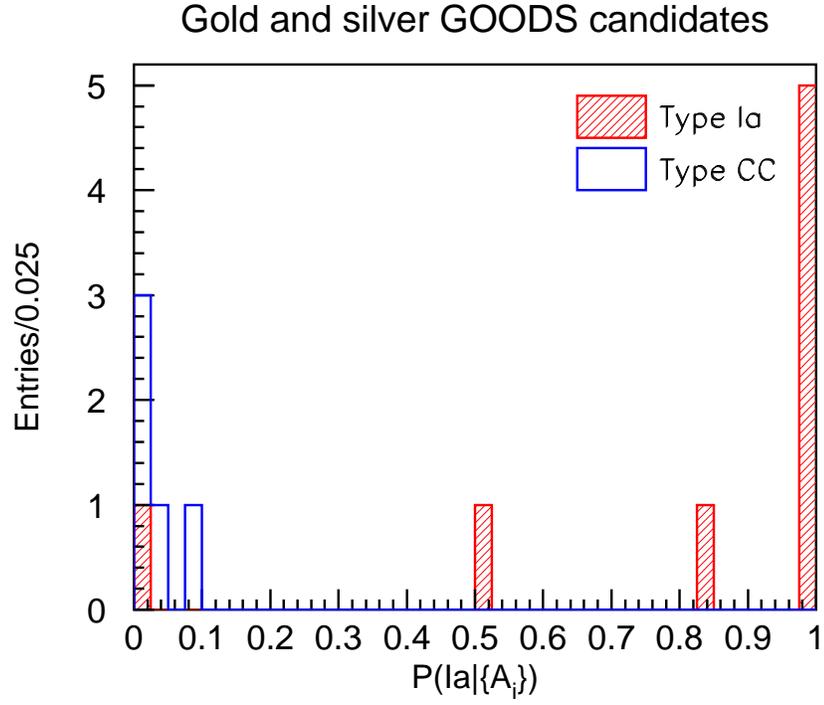}
  \end{center}
\caption[]
{\label{fig:space}
$P({\rm Ia}| \{A_i\})$ for the gold and silver 
GOODS candidates: Ia (red, hatched histograms) and CC (blue, empty histograms).
}
\end{figure}

We require that there be at least three data points with the signal-to-noise ratio S/N greater than
 2, and that there be at least two data points with S/N greater than 3.
This eliminates the ``single-epoch'' candidates
2003es, 2003eq,  2002lg, 2003ak, 2003eu, 2003al, and  2002fx (Ia's),
and  2003et and 2003er (CC's).
  We also eliminate one silver Ia candidate, 2003lv, which appears to have a residual cosmic ray 
 contamination.
%~\footnote{If this method is to be used in a rates analysis, 
% one needs to be careful not to put quality cuts in the data
% without understanding their effect on the acceptances of 
% known candidates.  As a rule, if one does not expect a strong
% contamination from anomalous events, it is best to leave these type
% of quality cuts out of the analysis. 
% As for cuts to remove cosmic ray contamination, it can
% be assumed to affect all candidates equally.  Therefore,
% the correction to the rates due to discarded events from 
% cosmic rays is trivial provided one has some idea of the 
% fraction of anomalies in the sample.}.
Figure~\ref{fig:space} shows the probabilities $P({\rm Ia}| \{A_i\})$
for the remaining  gold and silver candidates.
\clearpage
%%%%%%%%%%%%%%%%
%%%%%%%%%%%%%
\clearpage
It is apparent that only one of the remaining type Ia candidates
has a very low $P({\rm Ia}| \{A_i\})$.
This candidate, 2003eb ($P({\rm Ia}| \{A_i\})$ = $2\times 10^{-4}$), appears to be more likely to  be a 
type IIn than a type Ia.  
Note also that a silver Ia candidate, 2002fy, has a somewhat marginal 
$P({\rm Ia}| \{A_i\})$ of $\sim$0.52, 
because it appears about equally likely to be a Ia and a Ibc.

%This candidate, 2002lg, that has only two
%data points in the $Z$-band, and one in the $I$-band.
%Note that this candidate was not used for the
%cosmological analysis of~\cite{riess}.

\begin{figure}[!htb]
\begin{center}
$\begin{array}{c@{\hspace{0.0in}}c}
\includegraphics[width=3.0in]{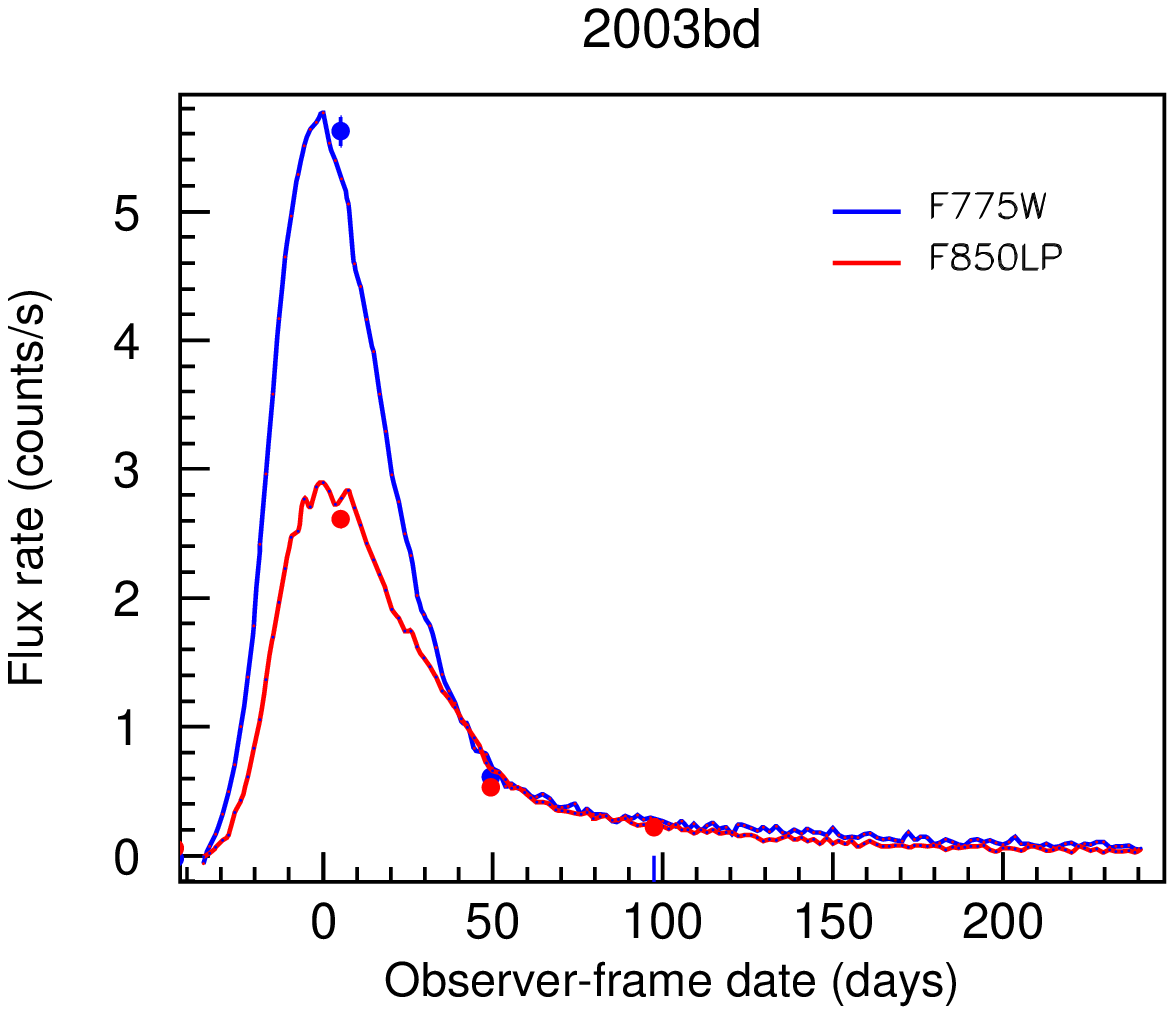} &
\includegraphics[width=3.0in]{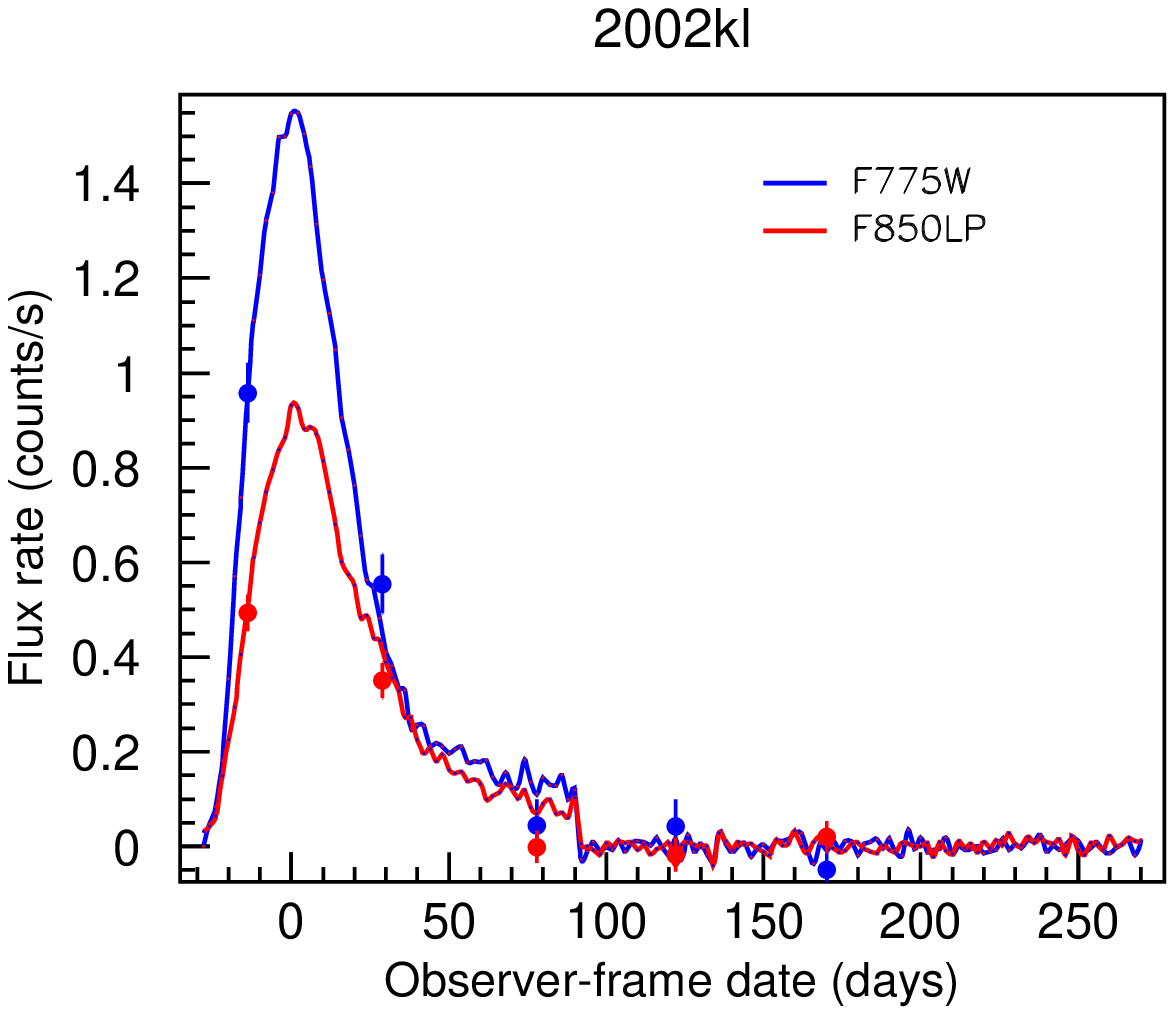}
\end{array}$
\end{center}
\caption{
The best-matching configuration for a gold type Ia candidate (2003bd) (left plot)
and for a silver type CC candidate (2002kl) (right plot).
Solid lines are the templates for type Ia (left plot) and type Ibc (right plot) supernovae;
filled circles with error bars are the data.   The discontinuity of the templates
in plot on the right is due to the fact that the available Ibc spectral template only
extends to 85 days in the supernova rest frame past the supernova's date of explosion.
}
\label{fig:best_space}
\end{figure}

When discussing these results, it is important to be clear that we are not performing a 
fit but calculating a single probability, where we sum over many configurations with
different stretches, magnitudes, \emph{etc.}
Nevertheless, it is useful  to consider the one configuration that best matches the data,
as this configuration will contribute the most to the final probability.
%best estimate of the magnitude, extinction,
%stretch (for Ia's), and shift by maximizing
%$P(P(m,s,R_v,A_v,Ia)|\{A_i\})$ as a function of these parameters
%to obtain $P_{max}(P(m,s,R_v,A_v,Ia)|\{A_i\})$;
%$P_{max}(m,s,R_v,A_v,Ia)|\{A_i\})$ is then the
%term that gives the largest contribution to
%the probabliity when marginalizing these parameters.
Looking at these ``best-matching'' configurations
is a good sanity check that the method is working.  
Figure~\ref{fig:best_space} shows such best-matching configurations 
%the light curves corresponding to $P_{max}(m,s,R_v,A_v, Ia)|\{A_i\})$
for two candidates, one Ia (2003bd) and one CC (2002kl).
The best match for the former is a Ia model,
while the best match for the latter is a Ibc model.
It is apparent that the best matches describe the observed data well.
\clearpage
%%%%%%%%%%%%%%%%%%%%%
%%%%%%%%%%%%%%%%%%%%%
\clearpage

%%%%%%%%%
\subsection{Monte Carlo Simulations of the GOODS sample}
\label{sec:mcspace}
%%%%%%%%%%%%
It is encouraging to see that even though the data sampling for these space-based data is very poor,
the method appears able to discriminate between different supernova types most of the time.
In order to better understand the performance of the method on such data, we create Monte 
Carlo samples 
of all of the supernova types considered using the same  sampling as that of the actual GOODS
data (5 epochs separated by $\sim$45 days).
We generate Monte Carlo samples of type Ia, Ibc, IIL, IIp, and IIn supernovae,
each with 500 candidates whose redshifts, 
stretches, magnitudes, and extinction parameters are selected randomly according to
the probability distributions in Eqns.\,(\ref{eqn:final_Ia}) and (\ref{eqn:final_T}).
We simulate both ACS F850LP and F774W  bands and 
impose the same signal-to-noise requirements on these simulated data as those
we chose for the GOODS data, as described below.

\begin{figure}[!htb]
\begin{center}
$\begin{array}{c@{\hspace{0.0in}}c}
\includegraphics[width=2.5in]{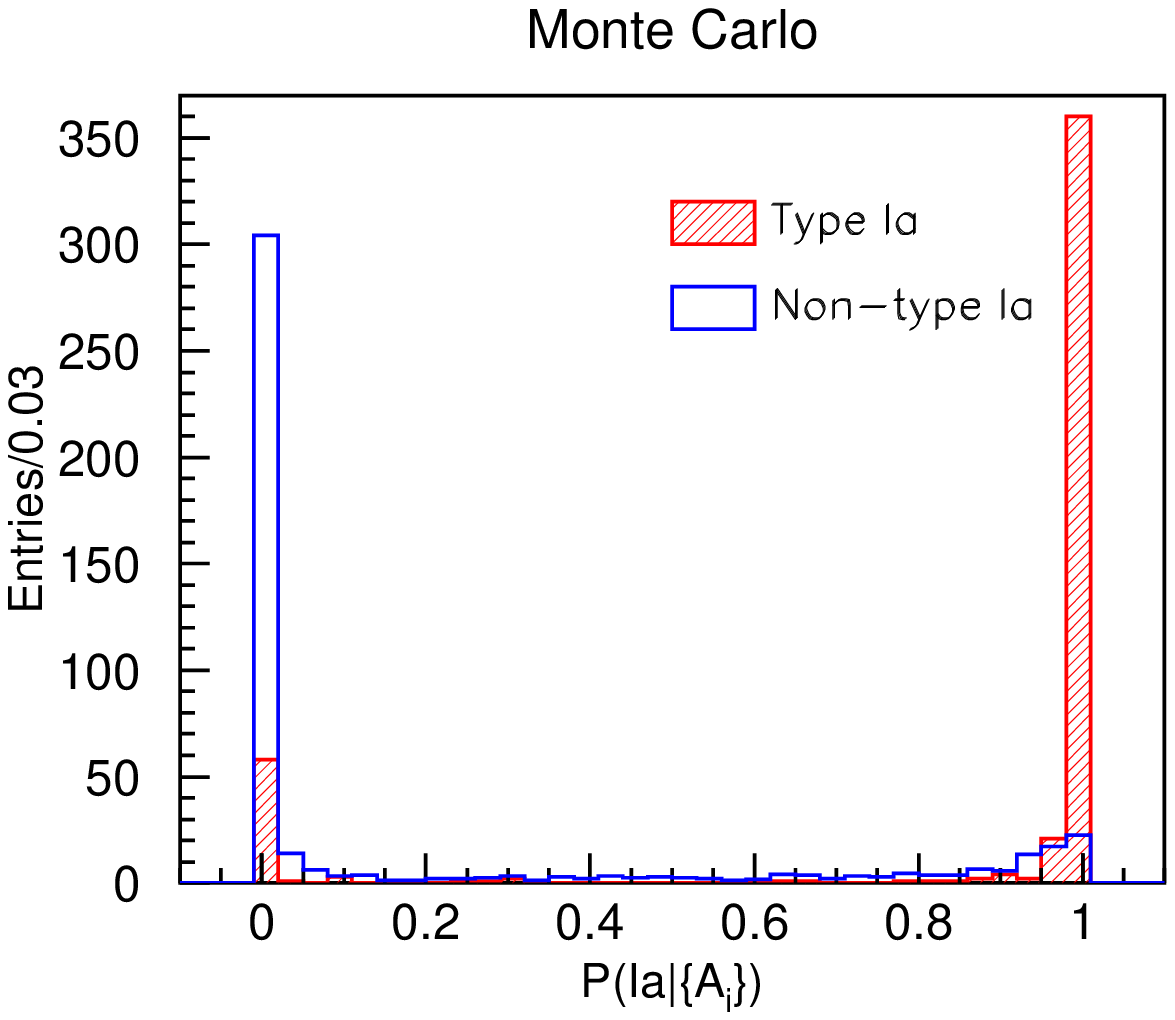} &
\includegraphics[width=2.5in]{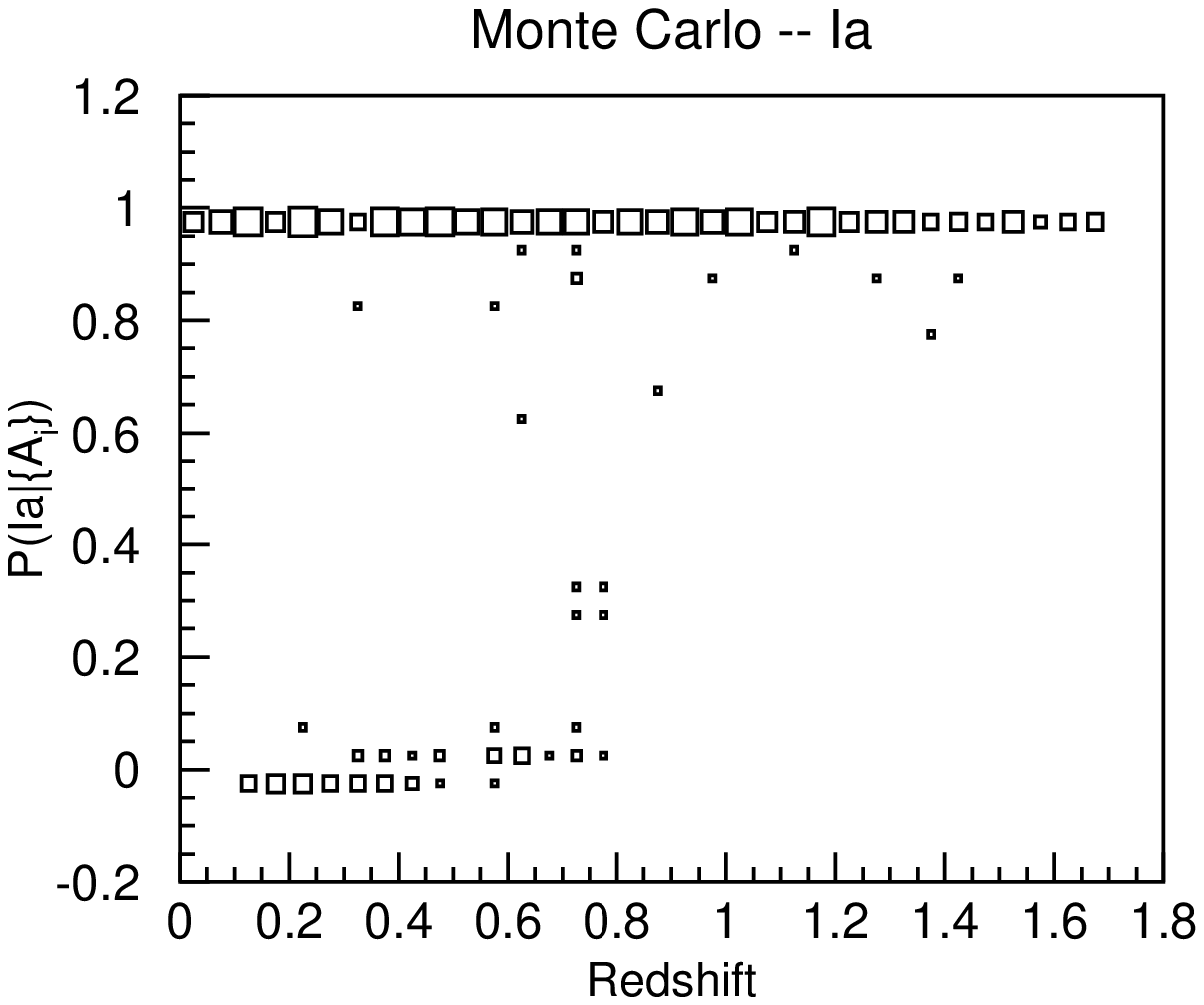} \\
\includegraphics[width=2.5in]{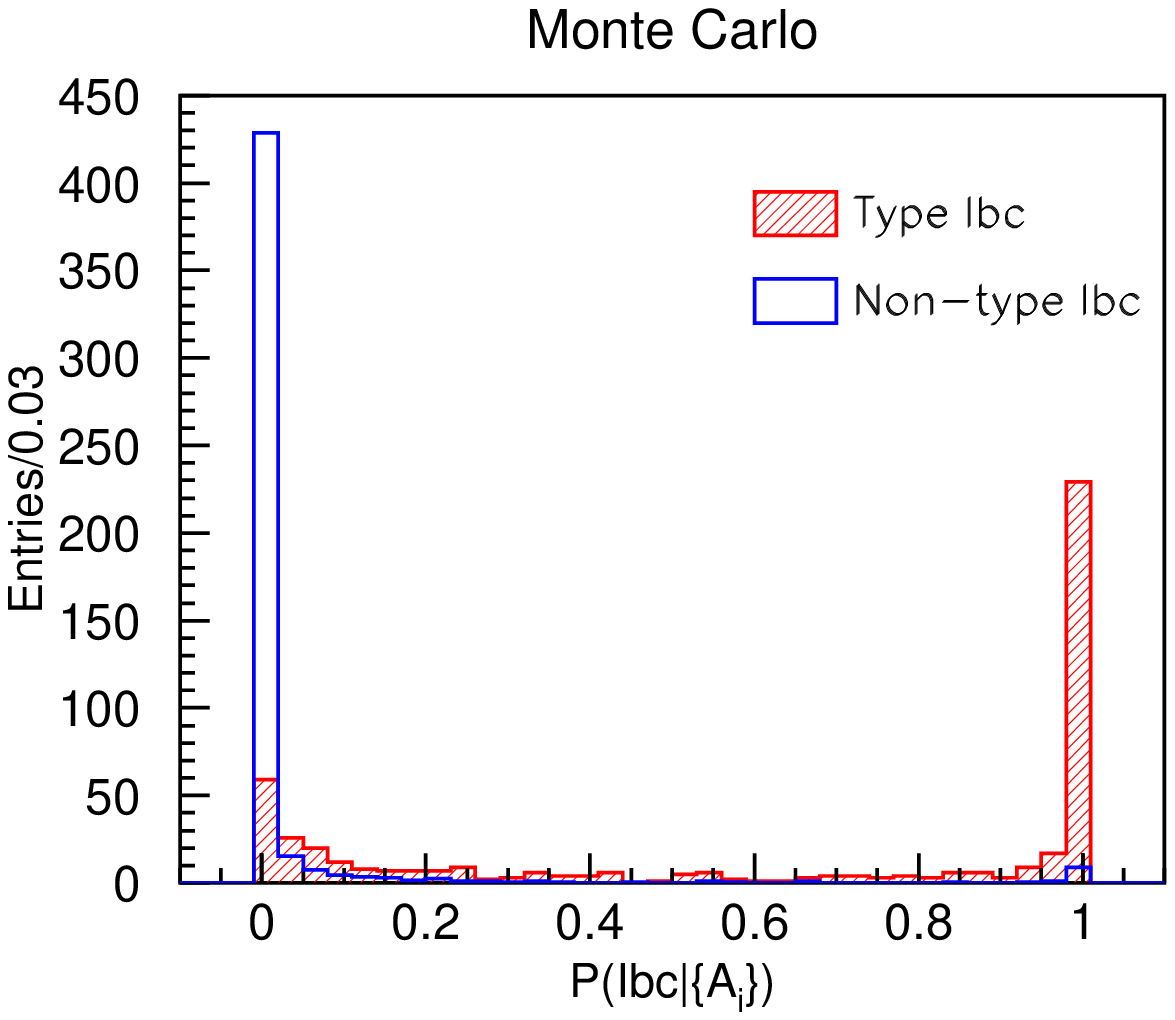} &
\includegraphics[width=2.5in]{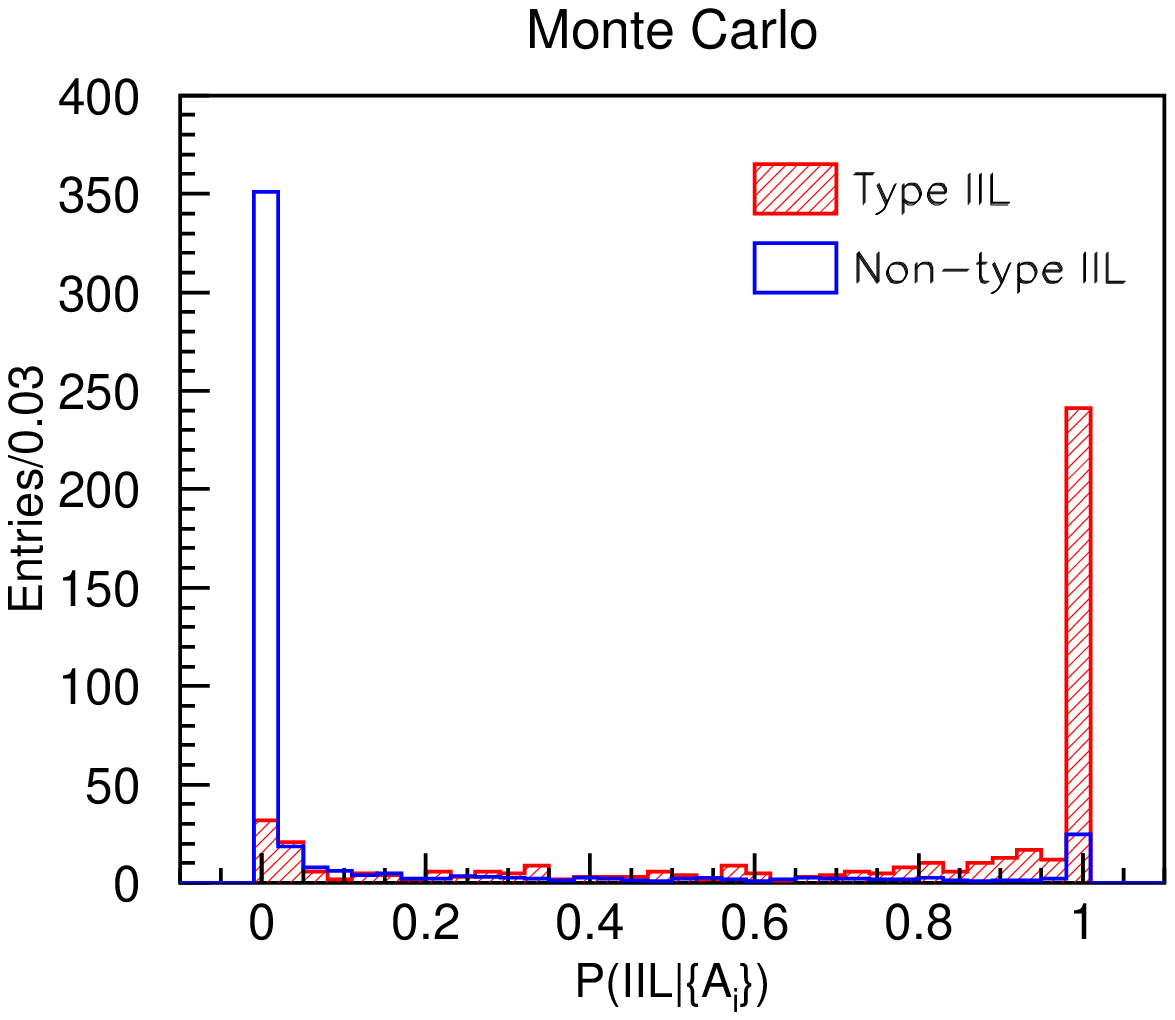} \\
\includegraphics[width=2.5in]{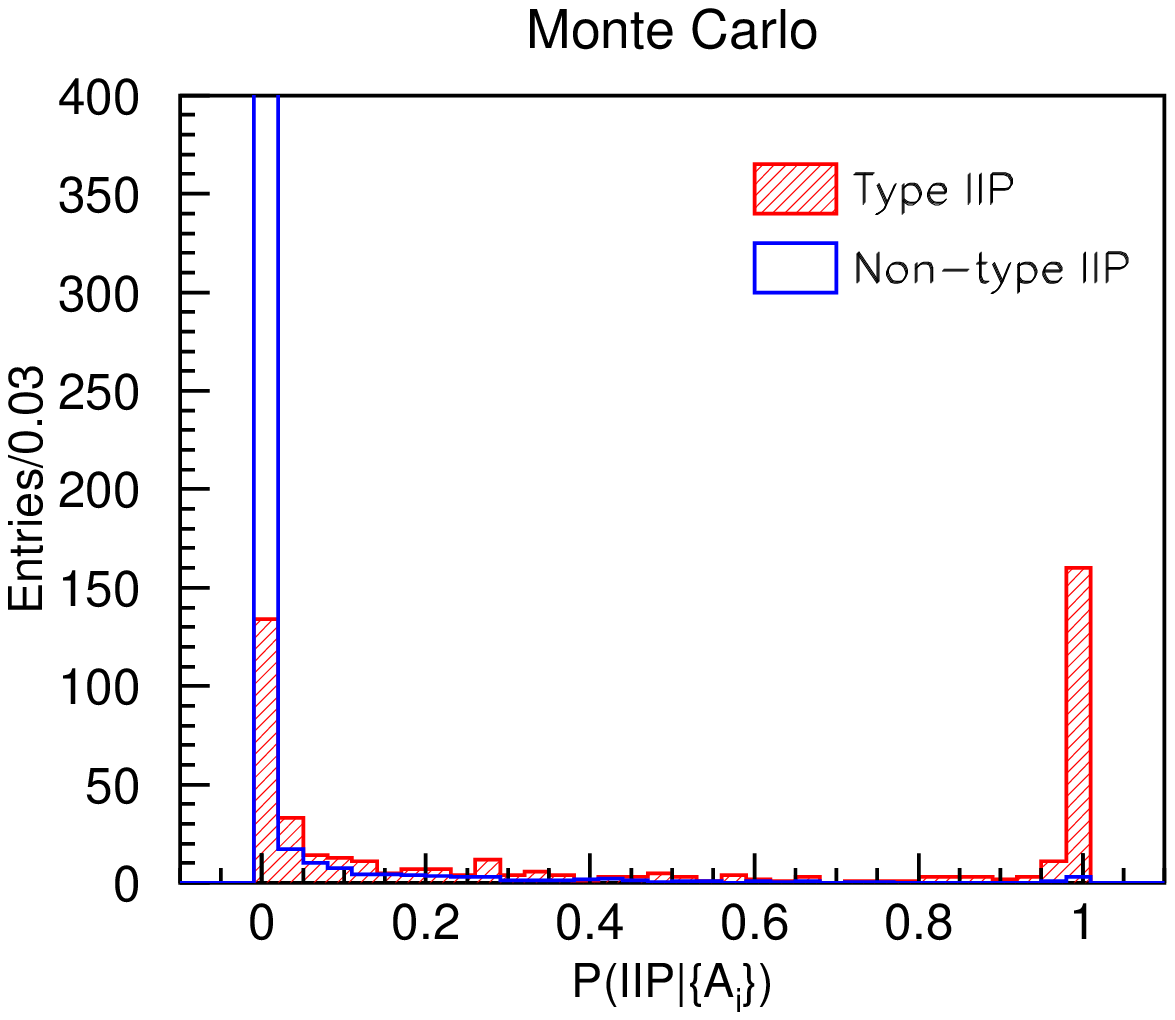} & 
\includegraphics[width=2.5in]{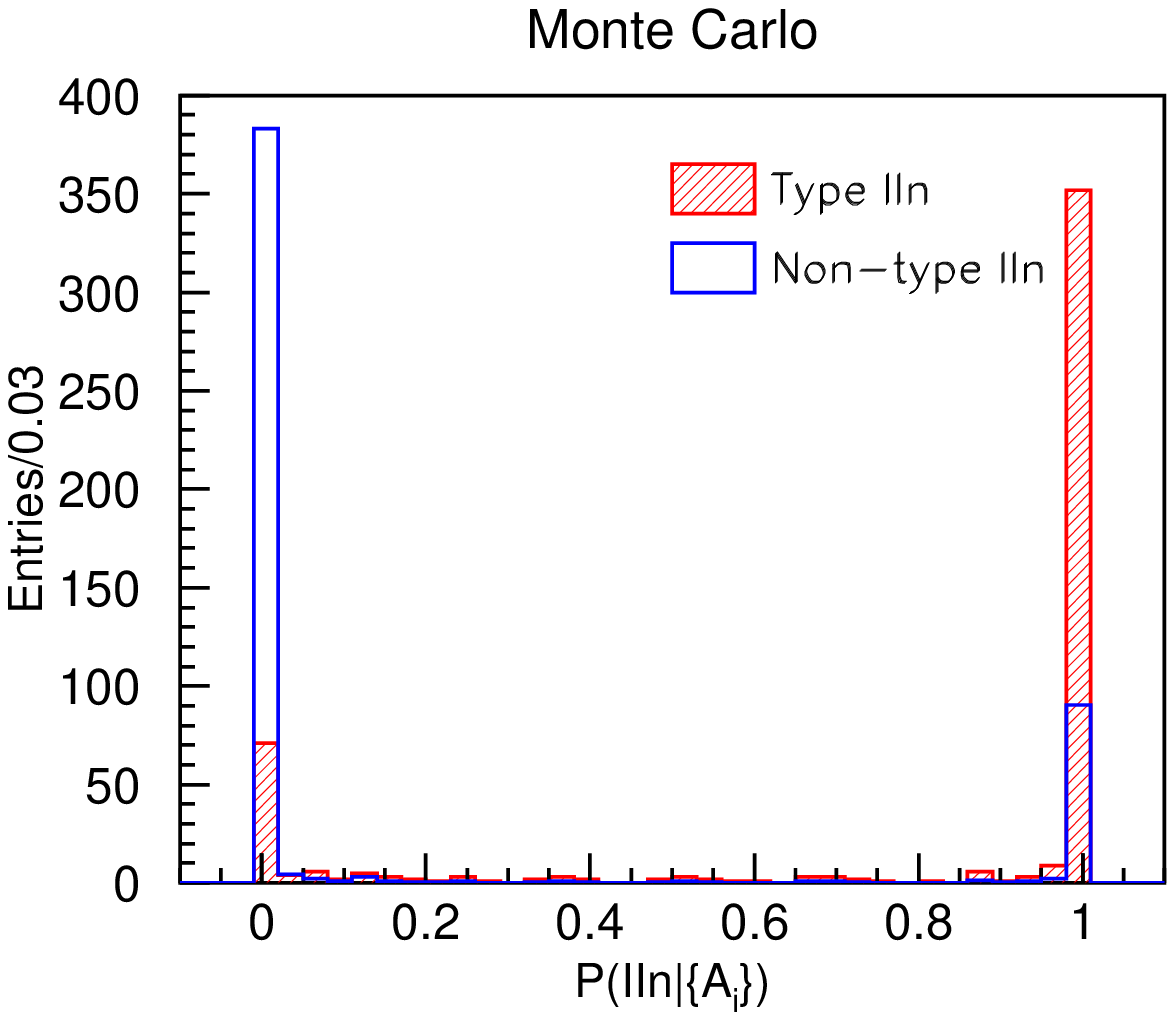} \\
\end{array}$
\end{center}
\caption{All but uppermost right: $P(T| \{A_i\})$ for Monte Carlo generated events.
The hatched histograms are for (from left to right, top to bottom):
$T = {\rm Ia}$, Ibc, IIL, IIP, and IIn, and the empty histograms
are for all other types $T' \ne T$.  
In all of the plots, the hatched and empty histograms are normalized to the area of the hatched histograms.
The near-zero $P(T| \{A_i\})$ for type T come from the events where the data populate
the tails of the light curves, making type discrimination difficult.
The uppermost right plot shows  $P({\rm Ia}| \{A_i\})$ for the Ia sample as a function 
of redshift.  The box size is proportional to the number of events in the bin.
}
\label{fig:mc}
\end{figure}

Figure~\ref{fig:mc} shows the resulting probability distributions for identifying 
type Ia supernovae and, as an example of the usage of the method for identifying other types,
type Ibc, IIL, IIP, and IIn supernovae.  
The near-zero probabilities for type $T$ supernovae in the $P(T| \{A_i\})$ plots
come primarily from the events where the data populate the tails of the light curves
and the discrimination between different types becomes particularly difficult.
The uppermost right plot in Fig.~\ref{fig:mc} shows $P({\rm Ia}| \{A_i\})$ for the type Ia 
Monte Carlo sample as a function of redshift; it indicates that the lower probability events 
are also those with lower redshifts.  This may simply be a consequence of the fact that our passbands of 
choice, F775W and F850LP, provide the best coverage for higher redshift candidates.
\clearpage
%%%%%
%%%%%
\clearpage

\begin{figure}[!htb]
\begin{center}
$\begin{array}{c@{\hspace{0.0in}}c}
\includegraphics[width=2.5in]{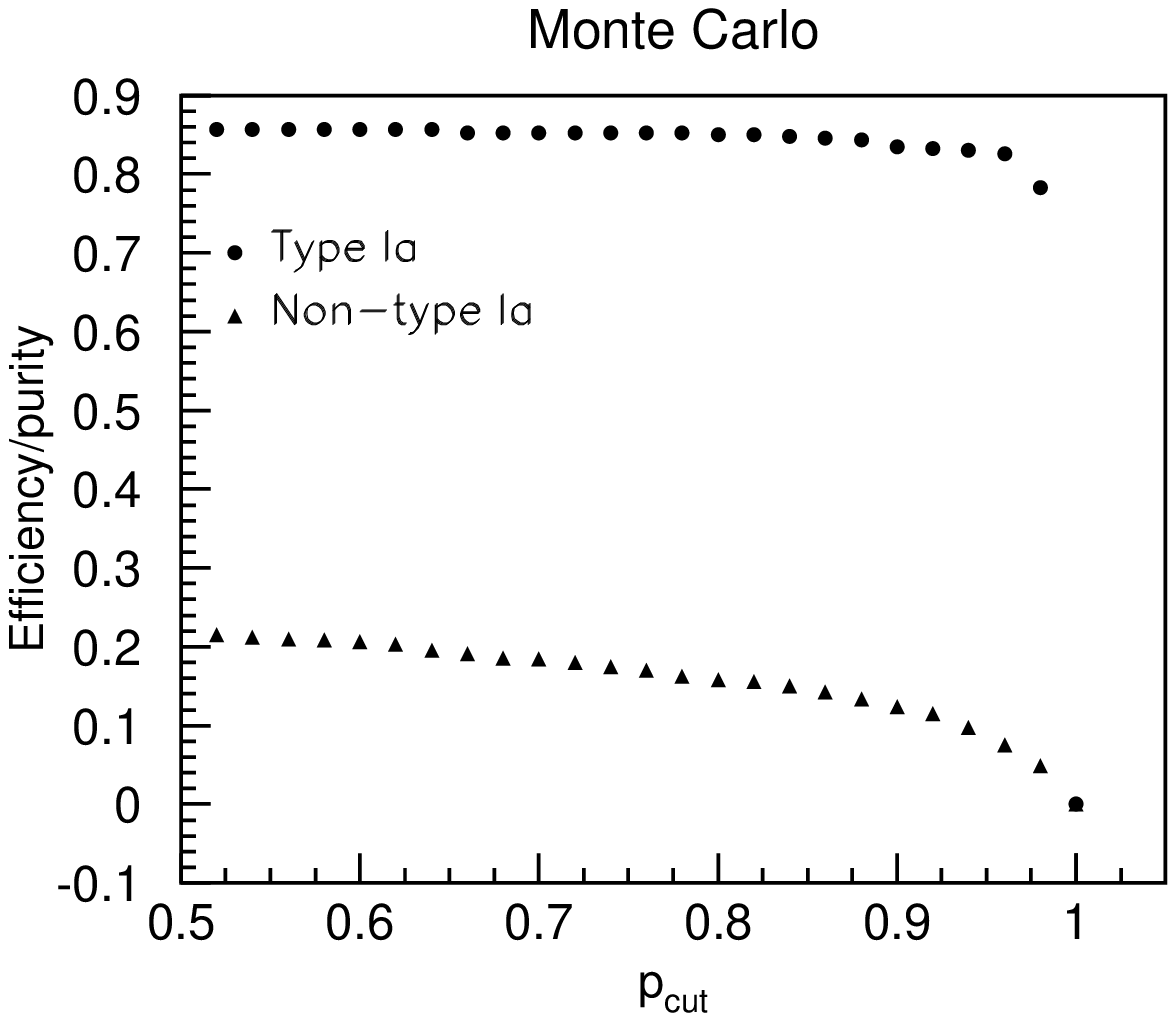} &
\includegraphics[width=2.5in]{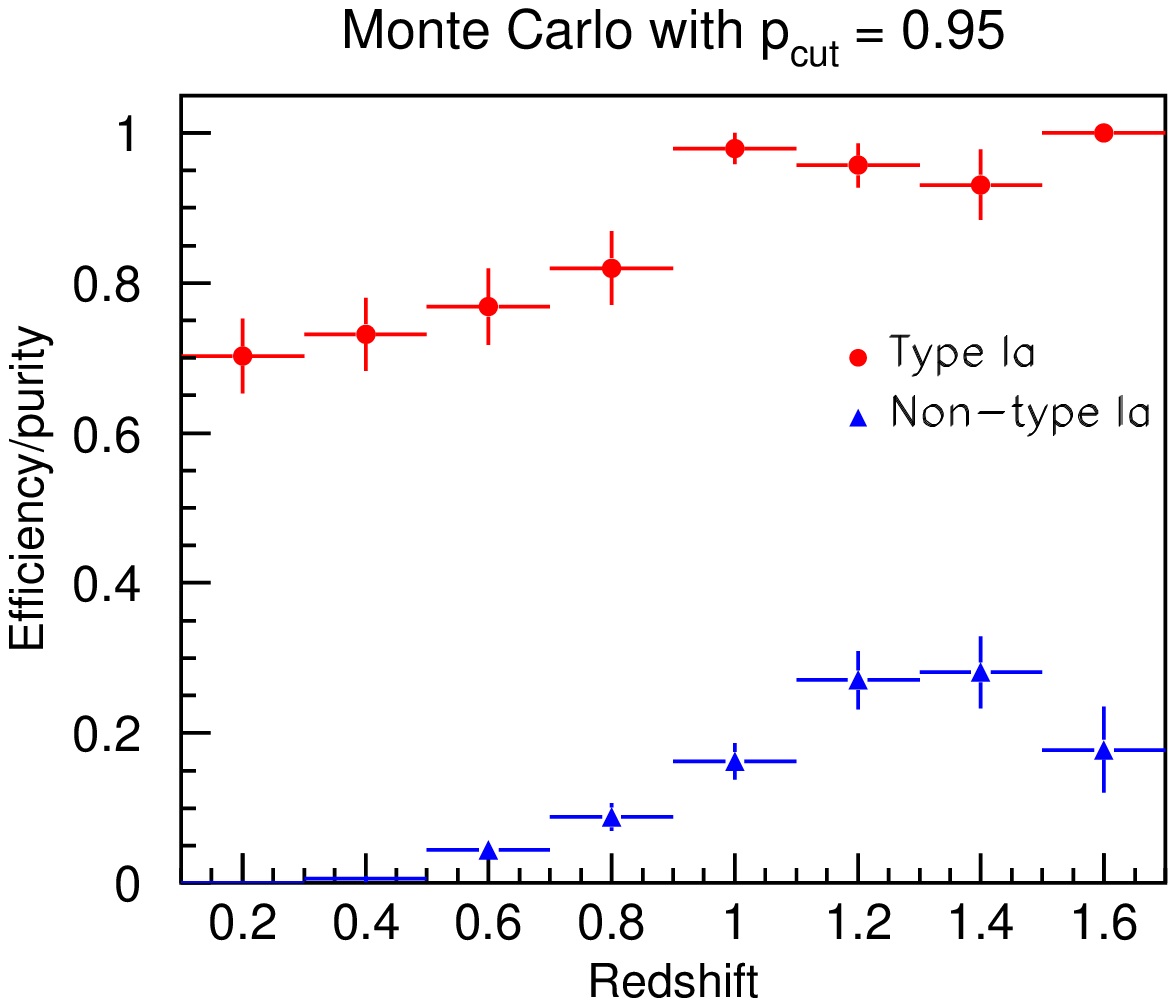} \\
\end{array}$
\end{center}
\caption{Left plot: the efficiency for type Ia simulated supernovae (filled circles) and purity 
for non-type Ia simulated supernovae (filled triangles) as a function of the cut $p_{\rm cut}$ 
on $P(T| \{A_i\})$ (require that $P(T| \{A_i\})$ $\geq$ $p_{\rm cut}$).  Right plot: 
the efficiency for type Ia simulated supernovae (filled circles) and purity 
for non-type Ia simulated supernovae (filled triangles) as a function of redshift 
for $p_{\rm cut}$ = 0.95.
}
\label{fig:mc1}
\end{figure}

To select a sample of candidates of a given type, one would apply a cut, $p_{\rm cut}$, 
on $P(T| \{A_i\})$, such that  $P(T| \{A_i\})$ $\geq$ $p_{\rm cut}$.
The choice of the cut of course affects both 
the efficiency (defined as the fraction of type Ia candidates
in a sample of Ia's that passed the cut) and the purity (defined here
as the fraction of non-type Ia candidates in a sample of non-Ia's that
passed the cut) of the sample.
Using our Monte Carlo samples, we show 
both the efficiency of selecting type Ia's
and the degree of contamination of the sample with non-type Ia's as a function
of $p_{\rm cut}$ in Fig.~\ref{fig:mc1} (left).   It is apparent
that the efficiency stays roughly flat, while the purity of the sample increases
dramatically, with higher $p_{\rm cut}$.  Figure~\ref{fig:mc1} (right) shows
the efficiency and purity distributions as a function of redshift for one choice of 
$p_{\rm cut}$, 0.95.  Similarly to the tendency exhibited in     
Fig.~\ref{fig:mc}(uppermost right), higher redshift candidates are both
the most efficiency identified and suffer the greatest degree of contamination.
\clearpage
%%%%
%%%%%
\clearpage

It should be emphasized that demonstrating the method using Monte Carlo samples
shows the behavior of $P(T|\{A_i\})$ in rather idealized circumstances; however,
it can and should be used for studying the method's performance.

%It is worth pointing out that even with well-sampled data consisting
%entirely of type Ia supernovae we do not expect to conclude that all of the
%candidates are Ia's, as there is always a very real possibility that 
%photometric data from Ia's can appear quite similar to that from other types.

%%%%%%%%%%%%%%%%%
\subsection{The Ground-based Data}
\label{sec:ground}
%%%%%%%%%%%%%%%%%
To demonstrate the method on well-sampled data, we use the Ia/Ia* candidates from the SNLS
collaboration~\citep{snls}.  Here, ``Ia'' denotes secure Ia's; and 
``Ia*'', probable Ia's, as defined in~\cite{howell}.
We generate the supernova templates 
in the four MegaCam bands where the data are available for most of the candidates: $g$, $r$,
$i$, and $z$.  Unlike the SALT fitter used by the SNLS collaboration~\citep{salt}, 
we do not restrict
ourselves to a particular restframe wavelength range for creating the templates 
(the acceptable wavelength range considered in SALT ranges from 3460 \AA \, to 6600 \AA).
The supernova templates may well be rather unreliable outside the chosen SALT range,
but we would like to find out to what extent they still offer type discrimination.

\begin{figure}[!htb]
\begin{center}
$\begin{array}{c@{\hspace{0.0in}}c}
\includegraphics[width=2.7in]{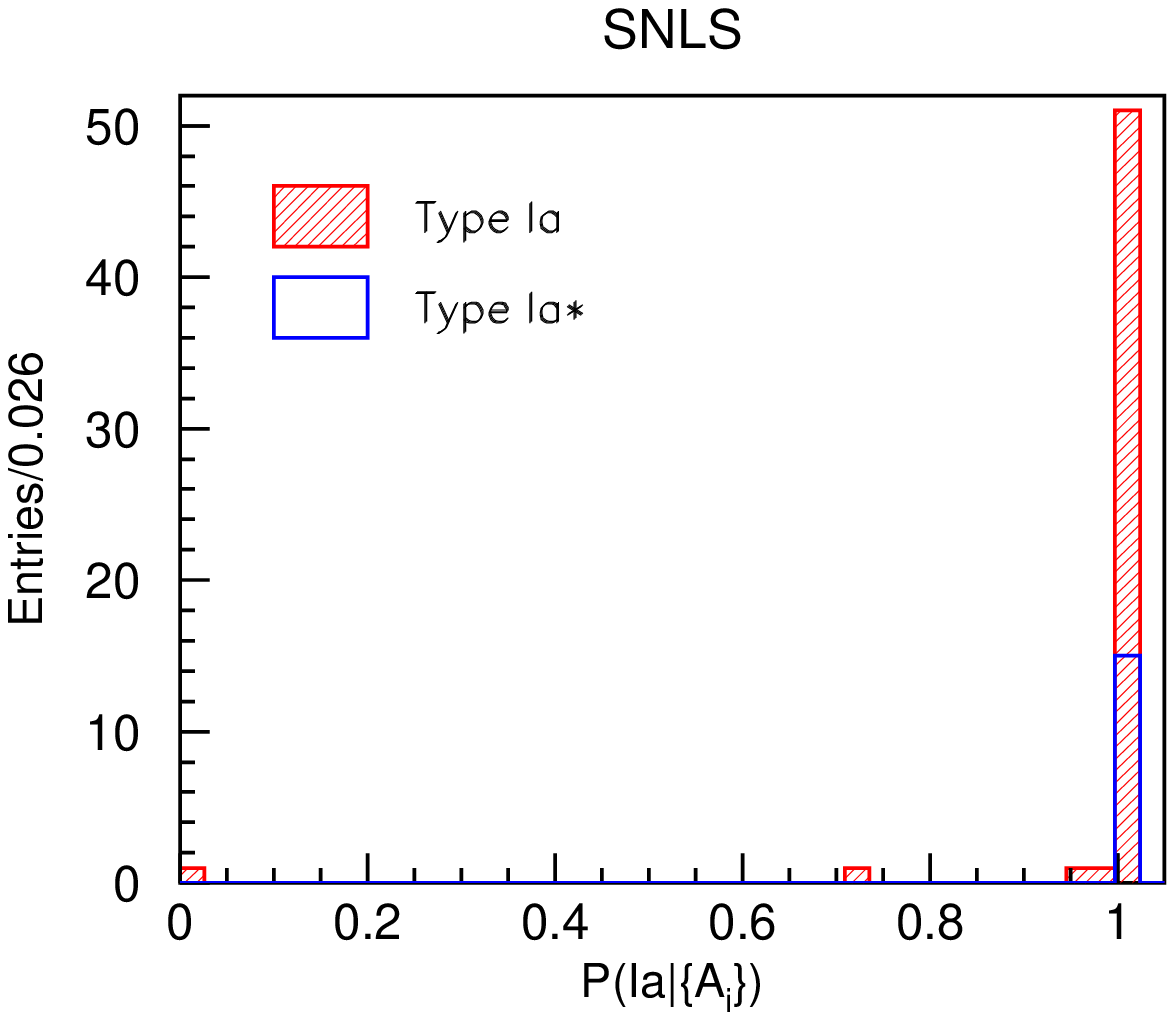} &
\includegraphics[width=2.7in]{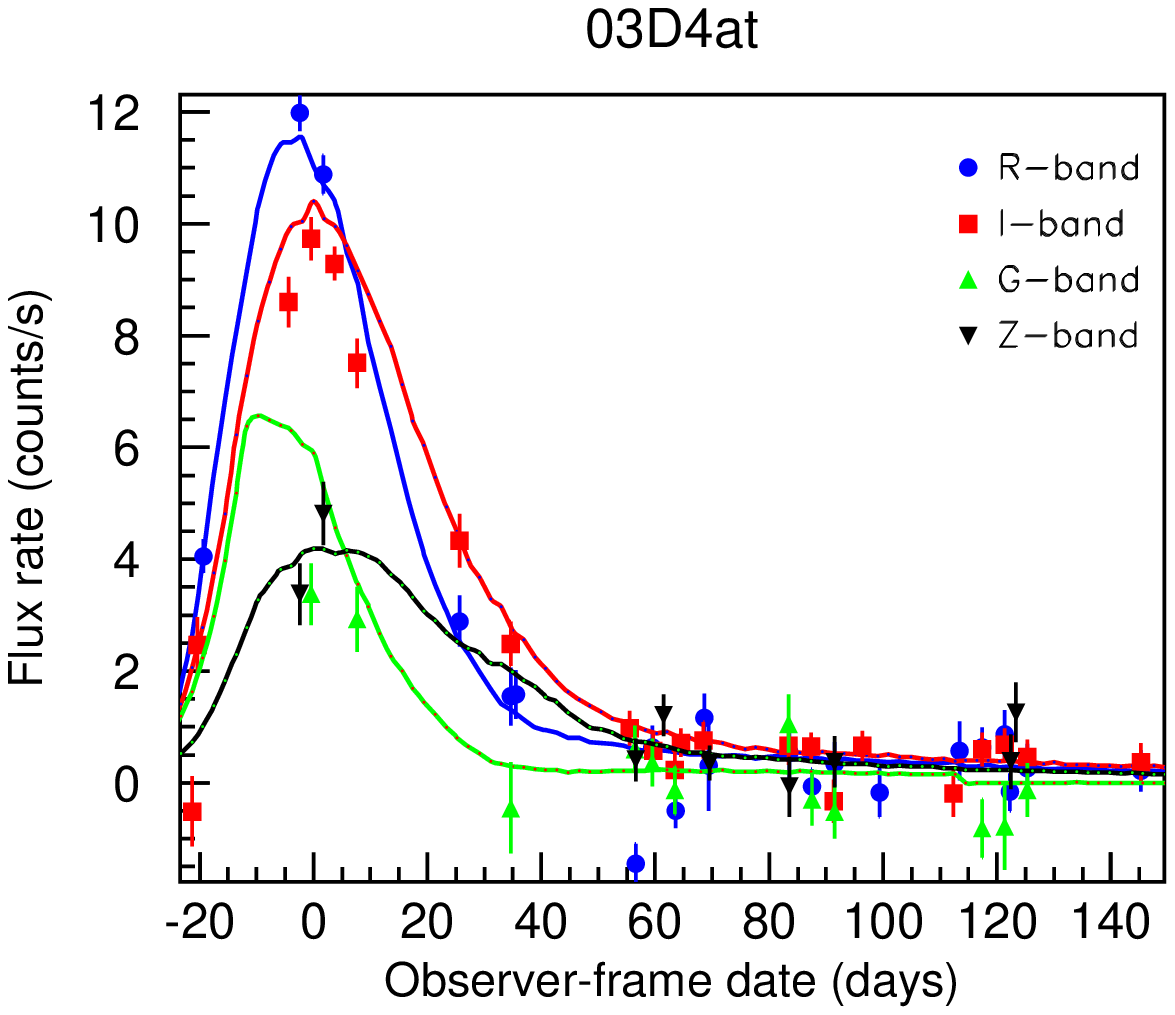} 
\end{array}$
\end{center}
\caption{Left plot: The probability distribution for Ia (hatched histogram) and Ia* (empty histogram) SNLS candidates.  
The plot does not show candidates 03D3bh, 03D4ag, and 04D3kr, which 
could not be identified as any known type (see Table~\ref{tab:badsnls}).
Right plot: 
The best-matching configuration for candidate 03D4at.  
Solid lines are the templates for a type Ia supernova; points with error bars are the data.
}
\label{fig:ground}
\end{figure}

Figure~\ref{fig:ground} (left) shows the probabilities for the 73 SNLS Ia/Ia* candidates to be Ia's,
and Fig.~\ref{fig:ground} (right) shows an example of the best-matching configuration
with respect to a type Ia model for one of the candidates, 03D4at.
\clearpage
%%%%
%%%%
\clearpage
Out of the 73 SNLS Ia/Ia* candidates, one had a probability that was lower than 0.5
and three were ``undefined'' 
(\emph{i.e.}, they yielded zero $P(T|\{A_i\})$ for all supernova types $T$ considered).
These four candidates are summarized in 
Table~\ref{tab:badsnls}.  
\clearpage
%%%%%%%
\begin{table}[htbp]
\begin{small}
  \begin{center}
    \begin{tabular}{|c|c|c|c|c|}
    \hline\hline
      Candidate   & Redshift   & $P({\rm Ia}|\{A_i\})$   &  SALT $P(\chi^2|{\rm DOF})$ & Comment\\
     \hline\hline 
%         03D4cn      & 0.82    & 0.397484  &  0.0396653  & Ibc almost as likely as Ia\\
%         03D1fq      & 0.80    & 0.0544602 &  0.399355  & Ibc more likely than Ia\\
         03D1gt      & 0.55    & 1.20e-9 & 0.86   & No $g$-band SALT fit \\
         03D3bh      & 0.25    & undefined   & 3.64e-17  & No $z$-band SALT fit\\
         03D4ag      & 0.28    & undefined   & 3.34e-27 & No $z$-band SALT fit\\
         04D3kr      & 0.34    & undefined   & 0  & No $z$-band SALT fit\\
     \hline\hline
   \end{tabular}
 \end{center}
\caption[]
{\label{tab:badsnls}
The four Ia/Ia* SNLS candidates that had low/undefined $P({\rm Ia}|\{A_i\})$.
The candidates with undefined probabilities are those which had zero 
$P(T|\{A_i\})$ for all supernova types $T$ considered.  Also listed
are the $\chi^2$ probabilities given  ${\rm DOF}$ degrees of
 freedom, $P(\chi^2|{\rm DOF})$, from the SALT fits.
 Note that for all of these candidates SALT did not fit one band ($g$ or $z$) 
 as the mean wavelength corresponding to the omitted band in the supernova restframe was
 outside the acceptable SALT range of [3460 \AA\, - 6600 \AA].
}
\end{small}
\end{table}
%%%%%%%%%
\clearpage
It should be noted that all four of these candidates had one band that was not fit
by SALT, as the mean wavelength corresponding to the omitted band in the supernova restframe was 
outside the acceptable SALT wavelength range.
Most other candidates in Fig.~\ref{fig:ground} (left) have a near-one  $P({\rm Ia}|\{A_i\})$.
Only one of them, 03D1fq, is somewhat of an outlier with a $P({\rm Ia}|\{A_i\})$ = 0.73.
This candidate has data in only two out of the four bands, $i$ and $z$.
% other ``outlier'' is 03D4cn  with p = 0.97; this one also has data in only two bands, i and z.
% for 03D4cn, (P(chi2, DOF) = 0.0396653
The four failures and the fact that for all of them there was at least one band
that was outside the [3460 \AA\, - 6600 \AA] restframe range may well signal 
a failure of the template to provide an adequate description of the supernova
behavior outside this range; however, the fact that only 4 out of 73 candidates failed
to yield desired discrimination is encouraging.

\begin{figure}[htb]
\begin{center}
$\begin{array}{c@{\hspace{0.0in}}c}
\includegraphics[width=2.7in]{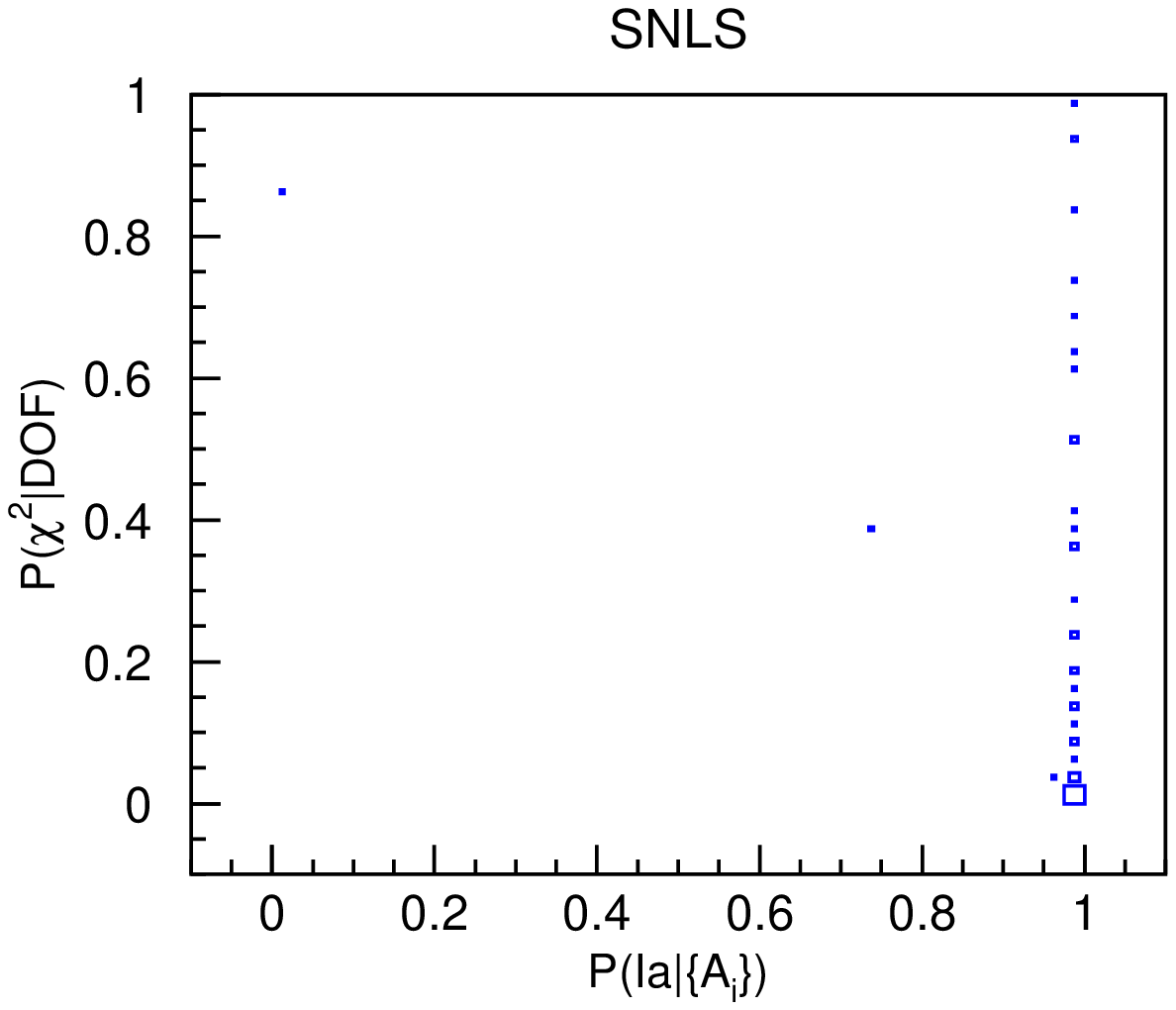} &
\includegraphics[width=2.7in]{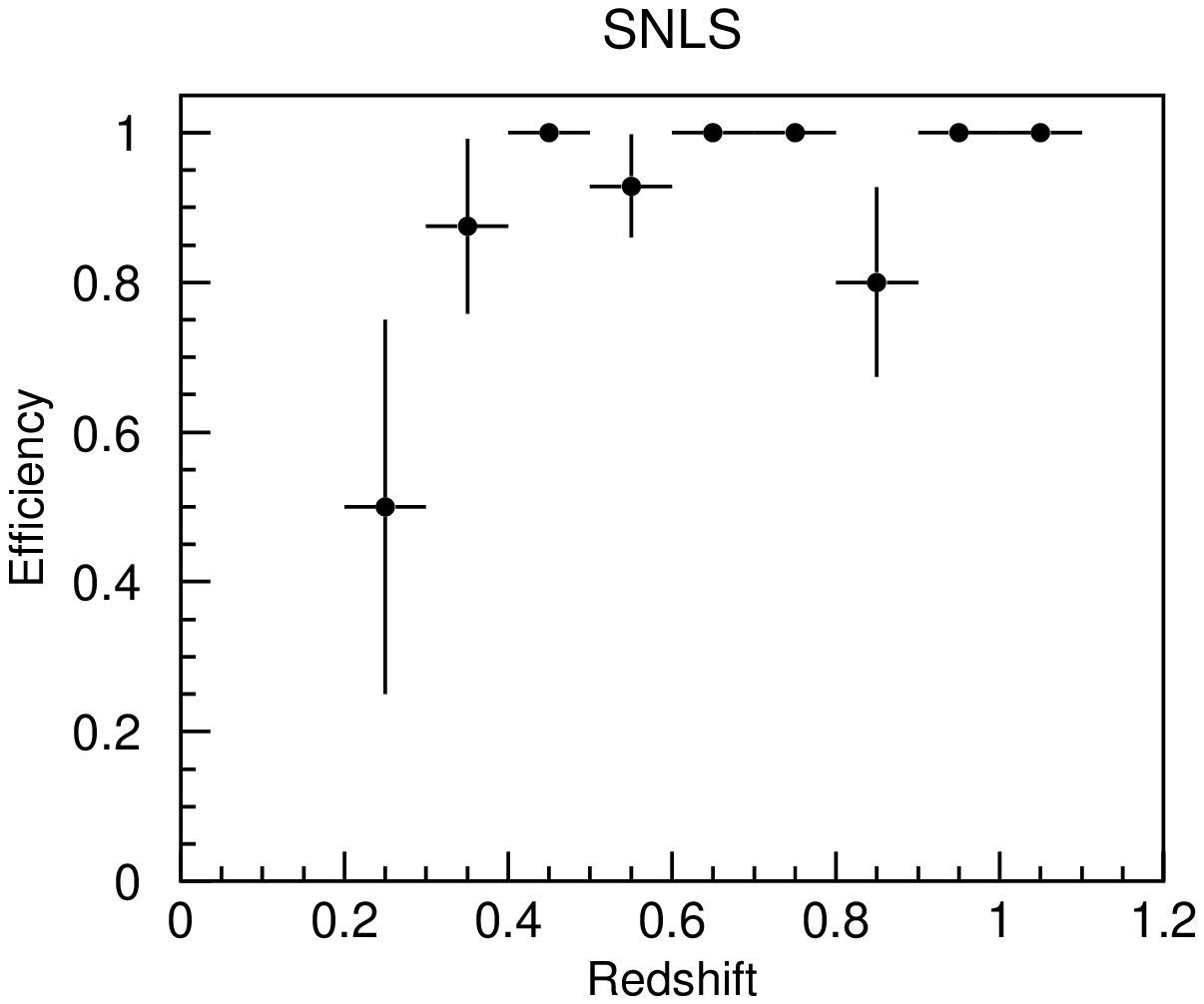} 
\end{array}$
\end{center}
\caption{
Left: The SALT fits $P(\chi^2|{\rm DOF})$ probability vs. $P({\rm Ia}|\{A_i\})$ for the 70 SNLS 
Ia/Ia* candidates (the figure excludes the three candidates with undefined $P({\rm Ia}|\{A_i\})$ 
listed in Table~\ref{tab:badsnls}).  The box size is proportional to the number of events in 
the bin.  Right: the efficiency of selecting type Ia supernovae for 
$p_{\rm cut}$ = 0.98 (we require $P({\rm Ia}| \{A_i\})$ $>$ $p_{\rm cut}$) as a function of redshift.
}
\label{fig:probvschi2}
\end{figure}

It is also interesting to compare our probabilities $P(T|\{A_i\})$ with the results
of the SALT fits -- namely, the $\chi^2$ probabilities given ${\rm DOF}$ degrees of
freedom, $P(\chi^2|{\rm DOF})$.  Table~\ref{tab:badsnls}
shows that the three ``undefined'' candidates also 
had exceptionally low $\chi^2$ probabilities from the SALT fits.
Figure~\ref{fig:probvschi2} (left) presents the $\chi^2$ probability of the SALT fits
vs. $P({\rm Ia}|\{A_i\})$.  One feature of this plot is the spread of the $\chi^2$ probabilities
for the candidates for which we calculated $P({\rm Ia}| \{A_i\})$ to be essentially one.
\clearpage
%%%%%%%%%%%%%%%%
%%%%%%%%%%%%%
\clearpage
Figure~\ref{fig:probvschi2} (right) shows the efficiency of selecting type Ia supernovae
in the SNLS sample for one choice of 
 $p_{\rm cut}$ = 0.98 (where we require $P({\rm Ia}|\{A_i\})$ to be greater than $p_{\rm cut}$), 
as a function of redshift.   It is apparent that the efficiency 
remains flat within errors.   We are unable to test the purity of this selection, as 
the non-type Ia candidates from the SNLS collaboration are not yet publicly available; 
however, we will certainly be able to do so when the data are released.

%%%%%%%%%%%%%%%%%
\subsection{Systematic Effects}
\label{sec:systematics}
%%%%%%%%%%%%%%%%%%
In order to investigate how changing the assumptions on the priors affects our results,
we performed a number of simple tests.  For all of the tests, we varied only one 
prior while keeping the rest of them unchanged.

\begin{itemize}
\item
Assuming that the stretch parameter for all type Ia candidates is 1 increases the number 
of type Ia/Ia* SNLS candidates that have low ($<$ 0.5) or undefined $P({\rm Ia}|\{A_i\})$  from 4 to 6.
It also lowers the probability for one gold Ia GOODS candidate, 2002hp, from 0.83 to 0.32; and for 
one silver Ia GOODS candidate, 2002fy, from 0.52 to 0.03.    

\item
Assuming no extinction increases the number of SNLS candidates with low/undefined
probabilities from 4 to 13.  This change does not significantly alter the probabilities
for the gold GOODS Ia candidates; however, it does 
lower $P({\rm Ia}|\{A_i\})$  for 2002fy to 0.15;
and increases $P({\rm Ia}|\{A_i\})$ for one silver CC candidate, 2002kl, from  0.01 to 0.80.

\item
Using a flat prior on the magnitudes does not change the number of undefined probability
SNLS candidates and moves candidate's 03D1fq probability from 0.73  to 0.43.
For the GOODS candidates, it lowers 
the probability for one silver Ia, 2002fy, from 0.52 to 0.14, but 
does not change the CC candidates' probabilities.

\item
The assumption of the flat prior on the supernova rates is undoubtedly incorrect;
however, it is difficult to suggest a plausible alternative since the rates 
for different supernova types (particularly for non-type Ias) are not well known, 
especially at high redshifts.  
Recent work~\citep[for example]{dahlen} seems to indicate that the
rates of CC supernovae with respect to type Ia's are higher by approximately
a factor of 3 at redshifts up to $z$ $\sim$ 1.   Assuming this ratio
holds at all redshifts, and modifying the prior accordingly, 
5 SNLS candidates end up with low or undefined probabilities.
As far as GOODS candidates are concerned, this change lowers $P({\rm Ia}|\{A_i\})$  for 2002fy 
from 0.52 to 0.27;    
but does not alter the distribution for the CC candidates.

\end{itemize}

%The choice of priors in any Bayesian method should be approached with care.
The choice of priors in any Bayesian method should be as complete as possible,
reflecting the  general belief about the 
relative probabilities of various choices of parameters.
For the data we have considered, the choice of the extinction prior appears to have the biggest effect,
although it is clear that all of the priors can have a dramatic effect on $P(Ia|\{A_i\})$
for some candidates.  
A change in any prior appears to be particularly consequential for 
the candidates whose classification was not strong to begin with (\emph{e.g.}, 
the starting $P({\rm Ia}|\{A_i\})$ with all of the default priors 
for the most affected gold Ia, 2002hp, was only about 0.83;
and for the most affected silver Ia, 2002fy, it was 0.52).

%%%%%%%%%%%%
\subsection{Further Improvements}
\label{sec:improve}
%%%%%%%%%%%%
While the method described in the paper appears to be quite promising, there are a number of 
improvements that are still to be explored.  
For instance, we have ignored the correlations between supernova
parameters by assuming the factorizable prior in Eqn.~\ref{equation:prior}.  Provided that these correlations are well known,
they can be included in a Hessian matrix within the Gaussians.  
Another improvement would be to use the Poisson distributions in lieu of 
the Gaussians in Eqn.~\ref{equation:likelihood}, especially for low photon statistics.  
In general, Poisson distributions have the added advantage of integrating
nicely to Gamma functions under certain circumstances.  
Here, the photon rates are average counts over time,
so they lend themselves better to Gaussian approximations.
However, one could
just as easily use absolute number of counts taken for a given 
epoch and use Poisson errors.   
Using better models, both for the supernova
behavior and for the behavior of the parameters that affect supernova
observations (\emph{e.g.}, the interstellar dust), is another issue that is 
important for providing a better discrimination.   The proposed method is of course
only as useful as the ability of the models it relies on to accurately represent
the actual diversity of the observed supernova sample; future ground- and 
space- based samples will yield tremendous improvements in this regard.

%%%%%%%%
\subsection{Purging Anomalies}
\label{sec:anomalies}
%%%%%%%%
The calculation of $P({\rm Ia}|{\rm candidate})$ is highly dependent 
on characterizing the full range of objects that may be in the candidate sample.
In practice, we must always consider the possibility that the candidate sample may be 
contaminated with ``anomalies'' that are merely mimicking a supernova signal -- \emph{e.g.}, 
certain known variable astronomical types, such as AGNs, or
residual ``objects'' resulting from poor image processing.
Furthermore, while there are now spectral templates for many special supernova types  (\emph{e.g.},
Ia 1991bg, 1991T, high velocity Ibc), there always exists the possibility that 
there are other, as of yet, undiscovered, supernova species.   
However, Eqns.~\ref{eqn:final_Ia} and~\ref{eqn:final_T} can be used to discard such 
anomalous supernovae with the following prescription.

First, we define a likelihood:
\begin{equation}
l(\vec{\theta},T|{\rm candidate})\, \equiv \, P(\{A_i\}|\vec{\theta},T)P(\vec{\theta},T)
\end{equation}
so that
%\begin{small}
\begin{equation}
l(T|{\rm candidate})\, = \, \sum_{\vec{\theta}} \, l(\vec{\theta},T|{\rm candidate}).
\end{equation}
%\end{small}
Then, for each candidate, we calculate 
the likelihoods for the various types $T$, $l(T|{\rm candidate})$, 
as described above.  We also generate large samples
of Monte Carlo events for each type considered (Ia, Ibc, IIL, \emph{etc.}),
and calculate a corresponding likelihood distribution \\
$l(T|{\rm simulated~type~T~candidate})$ 
 for each type $T$.
We define parameter $Q$ such that if
the probability of obtaining 
$l(T|{\rm simulated~type~T~candidate})$ $<$ $l(T|{\rm candidate})$ is
smaller than $Q$, then the candidate is designated an anomaly.
Effectively, $Q$ is the confidence level that the candidate is not a
type $T$ supernova.
Note that it is advantageous to consider each type $T$ in turn, rather
than the entire denominator of Eqn.~\ref{eqn:bayes}, since this approach does not force
one to make any assumptions about the relative numbers of supernovae
of type $T$ in the likelihood distributions.  

The parameter $Q$ must be larger than the inverse of the
number of simulated events for type $T$. In practice, the amount of CPU time limits 
the number of simulated supernovae that one can generate, and therefore constrains $Q$.
Note that this method does not remove events whose light curves are similar to known types;
$Q$ essentially defines how different the light curves of an anomaly must be from the known 
supernova light curves for it to be tagged an anomaly.  
Large values of $Q$ will possibly allow fewer anomalies that look similar to known supernova 
types to contaminate the ``purified'' sample, but will also 
decrease the statistics in the number of candidates of interest.  
A small $Q$ will require that anomalies differ more dramatically from 
the known supernova types, but it increases the statistics
for the candidates that we want to measure.  Small $Q$ 
values are also difficult to obtain as $Q$ is constrained
by the number of simulations.

Using parameter $Q$ to make a cut on $l(T|{\rm candidate})$ is quite 
similar to using the more traditional $\chi^2$ analysis for
candidate selection.  The $\chi^2$ probability quantifies how likely
it is that a measured distribution would have an equal or greater 
deviation from the model than the one observed.
Likewise, we calculate the likelihood $l(T|{\rm candidate})$ and
find how likely it is to for a particular candidate's data to fluctuate
to model $T$; if it is unlikely (\emph{i.e.}, the likelihood is 
less than $Q$) for all types, we call such a candidate an anomaly.

%We are, in effect, making a cut on $l(T|candidate)$ like 
%we would a $\chi^2$.   
%A $\chi^2$ is the probability that some curve deviates from some known
%distribution.  The $\chi^2$ probability that a distribution have a deviation
%greater than the one that is measured.  We define the likelihood
%$l(T|candidate)$, which is $P(T|candidate)$ without the normalization, and
%find how likely it is to fluctuate to that value.
%If it is unlikely (i.e. if the probability is less than $Q$),
%then it is called an anomaly.  In effect, humans do this procedure
%in  a less quantitative way by ``eye'', picking out candidates
%that do not look like anything they're seen before, or deviate
%dramatically from some known models.  

%Note that it might appear circular that we are using the same likelihoods for 
%rejecting anomalies as we will for typing the supernovae.  However, there is 
%no contradiction here, as we are simply relying on a 
%Monte Carlo simulation to establish the lower limit on the possible values of 
%the likelihoods for known supernova types.  

Note that this method would not effectively remove          
poorly sampled anomalies, as their probability would be relatively high
for every type.  The practicality of this approach for purging anomalies 
will be explored in a future paper.

Finally, note that using a method based on this simple likelihood criterion to 
provide the actual classification of supernovae is very different from our
Bayesian classification approach, since it does not account for 
the possibility that a number of supernova types might have the same values 
for $\vec{\theta}$.

%%%%%%%%%%%
\subsection{Fitting with the Maximum Likelihood}
\label{sec:fitting}
%%%%%%%%%%%
The Bayesian Adapted Template Match (BATM) method described in~\cite{barris}
uses a maximum likelihood fit to extract supernova parameters.  For completeness,
we also describe the procedure for maximizing the likelihood for a given 
parameter, as well as for estimating the errors on the parameter.

The formulation of $l(\vec{\theta},T|{\rm candidate})$ lends itself to a
maximum likelihood fit to
$t_{\rm diff}$, $M$, $s$ or $R_v$ and $A_v$ provided that the supernova type $T$ is known.
Indeed, we have maximized $l(\vec{\theta},T|{\rm candidate})$
to find the ``best-matching'' configurations
of these variables in Section~\ref{sec:valid} (see Figs.~\ref{fig:best_space}
and~\ref{fig:ground}(right plot)), where we see that it does an excellent job.
However, these configurations were found by maximizing
all the parameters at the same time.  To find the best estimate
for one parameter, we must assume the possibility that the others
have some range of values - each with a corresponding probability.
That is, we need only maximize one parameter at a time
and marginalize the rest.
In this scheme, errors are estimated
by integrating the likelihood to obtain
the upper and lower bound of the
68\% confidence region.

For instance, suppose that we want the best estimate
for the magnitude, $M$.  We first marginalize all the parameters
in the likelihood to obtain $l(M,T|{\rm candidate})$.  Then, assuming
that maximizing $l(M,T|{\rm candidate})$ gives the best estimate for the
fitted parameters, we maximize $l(M,T|{\rm candidate})$ to obtain the best estimate for $M$, $M_{max}$.
With the best estimate calculated, we then find the uncertainties by satisfying
\begin{small}
\begin{equation}
0.16=\frac{\int_{0}^{M_{max}-\sigma_{M-}}l(M,T|{\rm candidate})dM}{\int_{0}^{\infty}l(M,T|{\rm candidate})dM}
=\frac{\int_{M_{max}+\sigma_{M+}}^\infty l(M,T|{\rm candidate})dM}{\int_{0}^{\infty}l(M,T|{\rm candidate})dM}
\end{equation}
\end{small}
where $M_{max}+\sigma_{M+}$ and $M_{max}-\sigma_{M-}$ are
the upper and lower bounds of the $68\%$ confidence region.
This procedure would then be repeated for all the parameters for which
we want estimates, by maximizing $l(s,T|{\rm candidate})$, $l(t_{\rm diff},T|{\rm candidate})$, \emph{etc.}
Computationally, it may be necessary to replace the integrals with 
finite sums up to values where
the terms of the likelihood are negligible, provided there is a single peak 
in the likelihood distribution.

Note that there are circumstances where maximizing $l(M,T|{\rm candidate})$ 
does not give the best estimate for the fitted parameters.  
As an example,~\cite{harr} argue that a correction to the parameters is needed 
for this to be the case.
This is one crucial reason why 
the performance of the likelihood always needs to be tested with Monte Carlo calculations.

%%%%%%%%%%%%%%%%%%%%%%%%%%%%%%%%%%%%%%%%%%%%
\section{Conclusion}
\label{sec:concl}
%%%%%%%%%%%%%%%%%%%%%%%%%%%%%%%%%%%%%%%%%%%%
We introduced a novel method to determine the probability 
that a supernova candidate is indeed a certain type of supernova 
sing photometric information alone.
The probability is derived using a Bayesian approach.
We have tested the method on 
both poorly sampled HST GOODS space-based data and 
well-populated SNLS ground-based light curves, with good results
even for data where very few epochs available.
While we considered primarily the application of this method to
identifying type Ia supernova candidates, it can of course be used to identify any other 
known supernova types (see Fig.~\ref{fig:mc}).  The results of these studies 
show the method to be promising.

We have assumed that the candidate sample consists entirely 
of supernovae.  The method naturally incorporates a number of possible 
hypotheses for the supernova types, allowing one to 
introduce prior knowledge of the probability distributions 
for various supernova parameters, as well as to marginalize
them.  Because it is crucial that the 
sample consist of candidates that have reliable models
(\emph{i.e.}, well understood supernova types such as Branch-normal Ia, 
Ibc, \emph{etc.}), we also proposed a possible approach 
for eliminating anomalous or less well understood candidates
from the data.  As more models for 
different supernova types become available,
more candidates will pass this preliminary selection
to be further considered for typing.

We have used the best currently available supernova models  
for all of the supernova types considered.
While aware of their limitations, 
we note that using these models already yields good discriminating power;
as better supernova models become available from the ongoing and
upcoming ground- and space- based supernova surveys, classifying supernovae
with this approach will become proportionally more reliable.

The method described in this paper could be useful for a number of studies.
An obvious application is selecting a high-purity sample of type Ia's
for further analysis -- \emph{e.g.}, creating a Hubble diagram
and extracting cosmology.  In order to do so, one would re-parametrize
the likelihood in terms of the distance modulus $\mu$ and maximize 
$l(\mu,z,{\rm Ia}|{\rm candidate})$ to obtain the best estimates of 
the distance modulus and the redshift of the candidate, as well as the errors
on these parameters, as described in Section~\ref{sec:fitting}.
A pure sample of type $T$ supernovae could also be used for calculating the rates
of this class of objects, provided one performs careful 
Monte Carlo studies of the method's rejection factor for type $T$ supernovae.
Another potentially useful application would be adapting the method
for early-time light curve parameters in such a way as to make it a trigger
for type Ia supernovae in large supernova surveys.  
Other possible extensions of the method would include introducing prior information on 
supernova rates for the different supernova types, as well as 
building a Bayes factor to measure the relative likelihood
that a given candidate is more or less likely to be a Ia supernova than another
type.

%%%%%%%%%%%%%%%%%%%%%%%%%%%
\acknowledgements
%%%%%%%%%%%%%%%%%%%%%%%%%%
We are grateful to the anonymous referee for many useful comments and suggestions.
We would like to thank Reynald Pain for providing us with the SNLS type Ia supernova
photometry, and Mark Sullivan for making his stretch data available to us.
We are also grateful to Natalie Roe, Alex Kim, Kyle Barbary, Greg Aldering, Saul Perlmutter,
Eric Linder, and Tony Spadafora for fruitful discussions.  
We would especially like to thank Segev BenZvi for his careful reading of the paper draft
and his many useful suggestions.
%The simulation used in this study, SNAPsim,
%was created by the SNAP simulation group.  The software used to obtain the photometric data for the HST GOODS supernova candidates was created by the SCP collaboration.
NK is partially supported by the 
Director, Office of Science, Department of Energy, under grant DE-AC02-05CH11231.

\appendix

%%%%%%%%%%%%%%%%%%%%%%%%%%%%%%%%%%%%%%%%%%%%
\section{ \label{APP:appa} Simulating Supernova Templates}
%%%%%%%%%%%%%%%%%%%%%%%%%%%%%%%%%%%%%%%%%%%%
In order to create the template supernova light curves in the available filter bands,
we do the following.  We start with a spectral template from~\cite{nugent}.  
The templates are made primarily from the publicly available supernova data
assembled in the SUSPECT database.\footnote{http://bruford.nhn.ou.edu/$\sim$suspect/index1.html}
We consider five supernova types: Ibc, IIL, IIP, IIn, and Branch-normal Ia.  The Branch-normal Ia template
is based on~\cite{nugentia}, with some additional features such as the 
smoothing out of the UV part.  This template is for a stretch 1 Ia.
The Ibc template is based on~\cite{levan}; the IIL, IIP, and IIn templates are 
based on~\cite{gill}, with a contribution from~\cite{baron} for IIP's and 
from~\cite{dicarlo} for IIn's.

Using the template for a given type, we compute the expected flux of the supernova at a given redshift, 
assuming the $\Lambda$CDM cosmology ($\Omega_{\Lambda}$ = 0.7, $\Omega_M$ = 0.3,
and $w$ = const = -1).   
The templates are created for 12 different values of the peak
$B$-band magnitudes, in the range of $\pm$ 3 $\delta M$, as
 listed in Table~\ref{tab:mags}.    For type Ia's, each 
template is also generated for 14 different values of stretch, ranging
from 0.6 to 1.3.  Three types of templates are made: one assuming 
no interstellar extinction, and two with extinction parametrized
by the Cardelli-Clayton-Mathis parameters
 $(A_v,R_v)$ = $(0.4,2.1)$ and $(0.4,3.1)$.

The simulation enables one to simulate either a ground- or a space- based observatory.
The generated supernova flux is convolved with the atmospheric transmission (for the ground-based case),
as well as relevant telescope, detector, and filter transmissions (for either ground- or space- based case).
After the observations are thus generated, they are realized by an 
aperture exposure time calculator, creating light curves in each of the relevant filter
bands.  The signal-to-noise $S/N$ ratio is computed as:
\begin{equation}
\frac{S}{N} =  \frac{R_{SN}\, f_{sig} \, t}{\sqrt{R_{SN} \,f_{sig}\, t + R_{sky}\,A\,t  + R_{d} \,n_p \,t + N_{read}^2 \,n_p \,n_{exp} + N_{\Delta}}}
\end{equation}
where $R_{SN}$ is the supernova source rate, $f_{sig}$ is the fraction of the source
flux in the aperture, $t$ is the exposure time, $R_{sky}$ is the sky rate,
$A$ is the effective seeing area,
 (computed taking into account the atmospheric seeing for the ground-based case,
the diffraction radius
calculated for a given filter central wavelength, the detector diffusion, and the detector 
pixel size),
$R_{d}$ is the detector dark current rate, 
$N_{read}$ is the detector read noise, 
$n_p$ is the number of pixels in the aperture, 
$n_{exp}$ is the number of exposures, and $N_{\Delta}$ is 
the flatfielding contribution from the detector inter-pixel sensitivity variations:
\begin{equation}
N_{\Delta} = \left ((f_{sig} \, R_{SN} + R_{sky}\,A +   R_{d} \,n_p) \times t\, \Delta \right)^2
\end{equation}
where $\Delta$ is the flatfielding error, taken to be 10$^{-4}$.

%%%%%%%%%%%%%%%%%%%%%%%%%%%%%%%%%%%%%%%%%%%
\section{The Stretch Parameter and Non-type Ia Supernovae}
\label{sec:appb}
%%%%%%%%%%%%%%%%%%%%%%%%%%%%%%%%%%%%%%%%%%%%
The stretch parameter is not defined for non-type Ia supernovae.  
However, the calculation of $P(T|{\rm candidate})$ necessitates that
it be formally introduced for all candidates, regardless of type.
If it was not included, the sum over the stretch in the numerator and denominator
of Eqn.~\ref{eqn:bayes}
would weight the non-type Ia's by a factor of $s_{max}-s_{min}$.

To show how the stretch might be introduced for non-type Ia supernovae,
we define $T^\prime$ as some generic non-type Ia supernova type,
and 
\begin{equation}
\vec{\theta}^\prime \equiv (t_{diff},M,A_v,R_v).
\end{equation}
we then calculate $\sum_{\vec{\theta}} P(\{A_i\}|\vec{\theta}^\prime,T^\prime)P(\vec{\theta}^\prime,T^\prime)$ imposing 
Eqn.\,\ref{equation:prior}:
\begin{eqnarray}
\sum_{\vec{\theta}} P(\{A_i\}|\vec{\theta},T^\prime)P(\vec{\theta},T^\prime) =\sum_{\vec{\theta}^\prime} \sum_{-\infty}^\infty P(\{A_i\}|\vec{\theta}^\prime,s,T^\prime)P(\vec{\theta}^\prime,s,T^\prime) \\ \nonumber 
=\sum_{\vec{\theta}^\prime} \sum_{-\infty}^\infty P(\{A_i\}|\vec{\theta}^\prime,s,T^\prime)P(\vec{\theta}^\prime|T^\prime)P(s|T^\prime)P(T^\prime) .
\label{eqn:almost_there}
\end{eqnarray} 
where, not specifying anything about the stretch parameter $s$, 
we allow it to take on any value between $-\infty$ and $+\infty$.

Then, for type $T^\prime$ supernovae, we require  that 
the likelihood, $P(\{A_i\}|\vec{\theta}^\prime,s,T^\prime)$ remain unchanged for any choice of $s$.  That is, 
\begin{equation}
P(\{A_i\}|\vec{\theta}^\prime,s,T^\prime)=P(\{A_i\}|\vec{\theta}^\prime,T^\prime).
\end{equation}
Therefore, Eqn.\,\ref{eqn:almost_there} becomes 
\begin{eqnarray}
\sum_{\vec{\theta}^\prime} \sum_{-\infty}^\infty P(\{A_i\}|\vec{\theta}^\prime,s,T^\prime)P(\vec{\theta}^\prime|T^\prime)P(s|T^\prime)P(T^\prime)\\ \nonumber
=\sum_{\vec{\theta}^\prime} P(\{A_i\}|\vec{\theta}^\prime,T^\prime)P(\vec{\theta}^\prime|T^\prime)P(T^\prime) \sum_{-\infty}^\infty P(s|T^\prime).
\end{eqnarray}
If $p(s|T^\prime)$ is the probability density such that $P(s|T^\prime)=p(s|T^\prime)\Delta s$,
then probability theory requires
\begin{equation}
\sum_{-\infty}^\infty p(s|T^\prime) \Delta s = 1.
\label{eqn:integrate_s}
\end{equation}

However, Eqn.\,\ref{eqn:bayes} remains the same if we set 
\begin{equation}
p(s|T^\prime)=\frac{1}{s_{max}-s_{min}};
\end{equation}
as well as replace the upper and lower bounds in Eqn.\,\ref{eqn:integrate_s} with $s_{max}$ and $s_{min}$, 
respectively;
and use the same limits of integration, $s_{min}$ and $s_{max}$,
for both the numerator and denominator.
This is the reason for the 
choice of stretch prior in Eqn.~\ref{eqn:spriornonia}.

%%%%%%%%%%%%%%%%%%%%%%%%%%%
%\section{References}


\begin{thebibliography} {99}
\expandafter\ifx\csname natexlab\endcsname\relax\def\natexlab#1{#1}\fi

\bibitem[{Aldering {et~al}(2006)}]{ald} G.~Aldering \emph{et al.}, preprint astro-ph/0607030 (2006)
\bibitem[{Astier {et~al.}(2006)}]{snls} P.~Astier \emph{et al.}, A\&A. 447, 31 (2006)
\bibitem[{Baron {et~al.}(2004)}]{baron} E.~Baron \emph{et al.}, ApJ 616, 91 (2004)
\bibitem[{Barris and Tonry(2004)}]{barris} B.~J.~Barris and J.~L.~Tonry, ApJ 613, L21 (2004)
\bibitem[{Bhat {et~al.}(1997)}]{harr} P.~Bhat, H.~Prosper, and S.~Snyder, Phys. Lett. B 407, 73, (1997)
\bibitem[{Branch {et~al.}(1993)}]{branch} D.~Branch, A.~Fisher, P.~Nugent, AJ 106, 2383 (1993)
\bibitem[{Cardelli {et~al}(1998)}]{ccm} J.~A.~Cardelli, G.~C.~Clayton, and J.~S.~Mathis, ApJ 329, L33 (1988)
\bibitem[{Dahlen {et~al.}(2004)}]{dahlen} T.~Dahlen {\it et al.}, ApJ 613, 189 (2004)
\bibitem[{Di Carlo {et~al.}(2002)}]{dicarlo} E.~Di Carlo {\it et al},  ApJ 573, 144 (2002).
\bibitem[{Dickinson  {et~al}(2003)}] {goods1} M.~Dickinson \emph{et al.}, in the proceedings of the ESO/USM Workshop ``The mass of Galaxies at Low and High Redshift'' (Venice, Italy, October 2001), eds. R. Bender and A. Renzini (2003)
\bibitem[{Fruchter and Hook(2002)}]{drizzle} A.~Fruchter and R.~N.~Hook, PASP 114, 144 (2002)
\bibitem[{Giavalisco {et~al}(2004)}]{goods0} M.~Giavalisco \emph{et al.}, ApJ 600, L93 (2004)
\bibitem[{Gilliland {et~al.}(1999)}]{gill} R.~L.~Gilliland, P.~E.~Nugent, M.~M.~Phillips, ApJ 521, 30 (1999)
\bibitem[{Guy {et~al.}(2005)}]{salt} J.~Guy \emph{et al.}, accepted for publication in A\&A, preprint astro-ph/0506583 (2005)
\bibitem[{Hatano {et~al}(1998)}]{hatano} K.~Hatano, D.~Branch, and J.~Deaton, ApJ 502, 177 (1998)
\bibitem[{Howell {et~al.}(2005)}]{howell} D.~A.~Howell \emph{et al.}, to be published in ApJ, (2005)
\bibitem[{Johnson and Crotts(2005)}]{john} B.~Johnson and A.~Crotts, submitted to AJ, preprint astro-ph/0511377 (2005)
\bibitem[{Levan {et~al.}(2005)}]{levan} A.~Levan \emph{et al.}, ApJ 624 880 (2005)
\bibitem[{Nobili {et~al}(2005)}]{nob} S.~Nobili \emph{et al.}, A\&A, 437, 789 (2005)
\bibitem[{Nugent(2006)}]{nugent} P.~E.~Nugent, http://supernova.lbl.gov/$\sim$nugent/nugent\_templates.html (2006)
\bibitem[{Nugent(2002)}]{nugentia} P.~E.~Nugent, A.~Kim, and S.~Perlmutter, PASP 114, 803 (2002)
\bibitem[{Patil {et~al}(2006)}]{dust1} M.~K.~Patil, S.~K.~Pandey, D.~K.~Sahu, A.~K.~Kembhavi, accepted for publication in A\$A, preprint astro-ph/0611369
\bibitem[{Perlmutter {et~al}(1997)}]{perl} S.~Perlmutter \emph{et al.}, ApJ 483, 565 (1997)
\bibitem[{Poznanski {et~al.}(2002)}]{pozn02} D.~Poznanski \emph{et al.}, PASP, 114, 833 (2002)
\bibitem[{Renzini  {et~al}(2001)}]{goods2} A.~Renzini \emph{et al.}, in the proceedings of the ESO/USM Workshop ``The mass of Galaxies at Low and High Redshift'' (Venice, Italy, October 2001), eds. R. Bender and A. Renzini (2003)
\bibitem[{Richardson {et~al}(2002)}]{rich} D.~Richardson \emph{et al.}, AJ 123, 745 (2002)
\bibitem[{Riess {et~al.}(2004a)}]{riess2004} A.~G.~Riess \emph{et al.}, ApJ 600, L163 (2004a)
\bibitem[{Riess {et~al}(2004b)}]{riess} A.~G.~Riess \emph{et al.}, ApJ 607, 665-687 (2004b)
\bibitem[{Sullivan {et~al.}(2005)}]{sull1} M.~Sullivan \emph{et al.}, accepted for publication in AJ, preprint astro-ph/0510857 (2006)
\bibitem[{Sullivan {et~al}(2006)}]{sull} M.~Sullivan \emph{et al.}, accepted for publication in ApJ, preprint astro-ph/0605455 (2006)
\bibitem[{Valencic {et~al}(2004)}]{dust2} L.~A.~Valencic, G.~C.~Clayton, K~D.~Gordon, ApJ 616, 912 (2004)

%%\bibitem{strolger}{L.~G.~Strolger \emph{et al.}, ApJ 613, 200 (2004).}
%\bibitem{pyraf}{P.~Greenfield \emph{et al.}, in the proceedings of Astronomical Data Analysis Software and Systems XII, October 2002, Baltimore, MD (2002).}
%\bibitem{lacosmics}{P.~G.~Van Dokkum, PASP 113, 1420 (2001).}




%%%%
\end{thebibliography}
\end{document}